\documentclass[usenatbib]{mn2e}
\usepackage{epsfig}
\usepackage{amsmath}
\usepackage{ulem}

\def\gsim{\mathrel{\raise0.35ex\hbox{$\scriptstyle >$}\kern-0.6em 
\lower0.40ex\hbox{{$\scriptstyle \sim$}}}}
\def\lsim{\mathrel{\raise0.35ex\hbox{$\scriptstyle <$}\kern-0.6em 
\lower0.40ex\hbox{{$\scriptstyle \sim$}}}}
\def\gs{\mathrel{\raise0.35ex\hbox{$\scriptstyle >$}\kern-0.6em 
\lower0.40ex\hbox{{$\scriptstyle \sim$}}}}
\def\ls{\mathrel{\raise0.35ex\hbox{$\scriptstyle <$}\kern-0.6em 
\lower0.40ex\hbox{{$\scriptstyle \sim$}}}}

\def\kms {{\,\rm km\,s^{-1}}}

\def\lesssim{\mathrel{\hbox{\rlap{\hbox{\lower4pt\hbox{$\sim$}}}\hbox{$<$}}}}
\def\gtrsim{\mathrel{\hbox{\rlap{\hbox{\lower4pt\hbox{$\sim$}}}\hbox{$>$}}}}

\date{\today}
\title[Large-Scale Structures at $z=0.83$]
{Spectroscopically Confirmed Large-Scale Structures Associated to a $z=0.83$ Cluster}

\author[Tanaka et al.]{
\parbox[t]{\textwidth}{
Masayuki Tanaka$^1$,
Tadayuki Kodama$^{2,3}$,
Nobuo Arimoto$^2$,
Ichi Tanaka$^4$,
}
\vspace*{6pt}\\
$^{1}$Department of Astronomy, School of Science, University of Tokyo, Tokyo 113--0033, Japan \\
$^{2}$National Astronomical Observatory of Japan, Mitaka, Tokyo 181--8588, Japan \\
$^{3}$European Southern Observatory, Karl-Schwarzschild-Str. 2, D-85748, Garching, Germany\\
$^{4}$Astronomical Institute, Tohoku University, Aoba-ku, Sendai 980--8578, Japan
}

\begin{document}

\maketitle

\begin{abstract}
We present a discovery of definitive large-scale structures around
RXJ0152.7--1352 at $z=0.83$ based on spectroscopic redshifts.
In our previous papers, we reported a photometric identification
of the large-scale structures at $z\sim0.8$.
A spectroscopic follow-up observation was carried out on 8 selected
regions covering the most prominent structures to confirm their
association to the main cluster.
In six out of the eight fields, a well isolated peak is identified in
the distribution of spectroscopic redshifts at or near the cluster
redshift.
This is strong evidence for the presence of large-scale structures
associated to the main cluster at $z=0.83$.
It seems that there are two large filaments of galaxies at $z\sim0.837$ and
$z\sim0.844$ crossing in this field.
We then investigate stellar populations of galaxies in the structures.
The composite spectra are constructed from a number of red
member galaxies on the colour-magnitude sequence.
We consider three representative environments -- cluster, group, and
field -- to investigate the environmental dependence of their star
formation histories.
We quantify the strengths of the 4000$\rm\AA$ break ($D_{4000}$)
and the H$\delta$ absorption features and compare them with model predictions.
The ``cluster'' red galaxies do not show any sign of on-going or recent
star formation activities and the passive evolution can
naturally link them to the present-day red sequence galaxies in the
{\it Sloan Digital Sky Survey}.
In contrast, the red galaxies in ``groups'' and in the ``field''
tend to show signs of remaining and/or recent star formation
activities characterized by weak [OII] emissions and/or strong H$\delta$
absorptions.
Our current data seem to favour a scenario that
star formation is truncated in a short time scale ($<$1Gyr).
This would imply that galaxy-galaxy interactions are responsible for
the truncation of star formation.
\end{abstract}

\begin{keywords}
galaxies: evolution ---
galaxies: clusters: individual RXJ0152.7$-$1357
\end{keywords}

%
%
%------------------------------------------------------------INTRODUCTION
\section{Introduction}

\label{sec:intro}
Distribution of galaxies in the local Universe is highly
non-uniform as known from various redshift surveys
(e.g. CfA redshift survey; \citealt{delapparent86},
2dF; \citealt{colless01}, and SDSS; \citealt{york00}).
Based on these extensive spectroscopic surveys, a map of the
local Universe is drawn with an unprecedented accuracy.
Galaxies tend to be clustered by gravity and form large
filamentary and clumpy structures with voids in between the
filaments.
At intersections of filaments, gravitationally bound concentrations of
galaxies, which we call clusters of galaxies, are often seen.
These filaments and galaxy concentrations are cast all over
the Universe forming network structures.

It is of great interest to know how such large scale structures
of galaxies have been built up in the history of the Universe.
The large-scale structures have been extensively explored
even at high redshifts based on various photometric/spectroscopic
surveys (e.g. \citealt{kodama01,kodama05,shimasaku03,gal04}).
These surveys suggest that filamentary structures of galaxies seem
to have appeared since an early epoch of the Universe.
However, it is not well known how these structures developed
with time and became matured to the present-day structures.

The cosmic large-scale structures inversely mean that galaxies 
live in various environments -- some galaxies live in low density
filaments and some live in dense clusters.
It is now a widely accepted fact that galaxy properties such as
star formation rates depend on environment that
surrounds galaxies (see \citealt{tanaka05} and references therein).
However, the origin of such strong environmental dependence of galaxy
properties is not well understood.
It must be closely related to the process of structure formation,
but physical mechanisms that actually drive the observed
environmental dependence remain unidentified.
To improve the situation, we are conducting a panoramic imaging and
spectroscopic surveys of distant clusters (PISCES; \citealt{kodama05})
from their cores
and to the surrounding filamentary structures.

During the course of this wide area survey, we have identified for the
first time the environment where star formation activity drops sharply.
That is ``groups of galaxies'' located in the filaments extending
out from the cluster cores.
\citet{kodama01} and \citet{tanaka05} have shown that the colour
distribution of galaxies changes at relatively low projected
density of galaxies that corresponds to groups of galaxies
in the filaments.
However, the actual physical process behind the truncation of
star formation is yet unidentified.
In order to further investigate what is happening in groups,
spectroscopic information is essential such as [OII]  emission and
Balmer absorption lines.
The spectroscopic information resolves the on-going star
formation activity and recent star formation history of the galaxies,
which cannot be resolved by colours
(e.g. \citealt{couch87,dressler92,balogh99,poggianti99}).

In this paper, we focus on one of the clusters from our PISCES sample,
RXJ0152.7--1357 at $z=0.83$, for which we already have a complete
optical data-set with Suprime-Cam \citep{miyazaki02}.
The cluster RXJ0152.7--1357 (RXJ0153 for short) is one of the most
X-ray luminous distant ($z>0.7$) clusters known.
It was discovered by various X-ray surveys \citep{rosati98,scharf97,ebeling00,romer00}.
Since these discoveries, the cluster has been investigated in various ways: X-ray
observations \citep{dellaceca00,maughan03}, the Sunyaev-Zeldovich effect \citep{joy01},
near-IR imaging \citep{ellis04}, and the weak lensing analysis \citep{huo04,jee05}.
\citet{demarco05}, \citet{homeier05}, and \citet{jorgensen05} presented
spectroscopic studies on this cluster.
Recently, \citet{girardi05} discussed dynamical properties of this cluster in detail.

We photometrically discovered large-scale structures around RXJ0153
based on the Subaru data as reported in \citet{kodama05}.
However, the photometric identification of the structures is subject to projection effects.
The discovered structures might be chance projections of galaxies along the line of sight.
To confirm the structures against the projection effects, 
a spectroscopic follow-up observation was carried out.
In this paper, we first present spectroscopic confirmation of the structures at $z\sim0.8$. 
We then discuss detailed star formation histories of galaxies from the spectra
and constrain physical processes that trigger the truncation of star formation.

The outline of this paper is as follows.
In section \ref{sec:obs}, we briefly review our photometric identification of
large-scale structures at $z\sim0.8$ around RXJ0153.
Our spectroscopic observations of RXJ0153 and data reduction are described as well.
Then we present results of the spectroscopic observation in section \ref{sec:lss}, 
and we address star formation histories of galaxies in section \ref{sec:sfh}.
Implications of our results are discussed in section \ref{sec:discussion}.
Finally, the paper is summarized in section \ref{sec:conclusion}.

Throughout this paper, we assume a flat Universe of
$\Omega_{\rm M}=0.3,\ \Omega_{\rm \Lambda}=0.7$ and $H_0=70\kms \rm Mpc^{-1}$.
Magnitudes are on the AB system.

%
%
%------------------------------------------------------------OBSERVATION AND DATA REDUCTION
\section{Observation and Data Reduction}
\label{sec:obs}

%-----------------------IMAGING OBSERVATION
\subsection{Photometric Identification of Large-Scale Structures at $z\sim0.8$}

We briefly review the photometric identification of large-scale structures
at $z\sim0.8$. Readers are refereed to \citet{kodama05} for details.

We observed RXJ0153 with Suprime-Cam in $VRi'z'$ under excellent conditions.
The seeing was $\sim0''.6$ in all bands.
Since the cluster lie at a high redshift and our imaging is deep (we can reach
$M^*_V+4$ galaxies at the cluster redshift), galaxies at the cluster redshift are
heavily contaminated by fore-/background galaxies.
To reduce the contamination, we applied the photometric redshift code of
\citet{kodama99} and extracted galaxies around the cluster redshift.
The distribution of photo-$z$ selected galaxies revealed a highly disturbed shape of RXJ0153
-- the cluster is elongated along the NE--SW direction and it accompanies clumpy subgroups.
In particular, the wide field of view (abbreviated as FoV in what follows) of
the Suprime-Cam has enabled us to discover a number of groups/clusters
surrounding the central cluster.

However, it is unclear if these groups and clusters are physically associated to
the central cluster due to the wide redshift slice adopted ($0.78<z_{phot}<0.86$).
This wide range was chosen so that we do not miss galaxies at the cluster redshifts
due to the error in the photometric redshifts.
The discovered systems could lie at different redshifts within the above redshift range.
Moreover, the projection effect is a concern.
Some of these systems could be chance projections of galaxies
along the line of sight.
Therefore, more accurate redshift measurements based on spectroscopic
observations are essential to identify real physical associations
to the main body of the cluster.

%-----------------------SPECTROSCOPIC OBSERVATION
\subsection{Spectroscopic Observation}
\label{sec:spec_obs}
We conducted a spectroscopic follow-up observation during 11--14 October 2004
with FOCAS \citep{kashikawa02} in MOS mode.
We used a 300 lines $\rm mm^{-1}$ grating blazed at 5500 $\rm \AA$
with the order-cut filter SY47.
The wavelength coverage was between 4700$\rm\AA$ and 9400 $\rm\AA$ with a pixel
resolution of $1.4\rm \AA\  pixel^{-1}$.
A slit width was set to $0\farcs 8$, which gave a resolution of
$\lambda/\Delta\lambda\sim500$.

\begin{figure*}
\begin{center}
\leavevmode
\epsfxsize 1.0\hsize \epsfbox{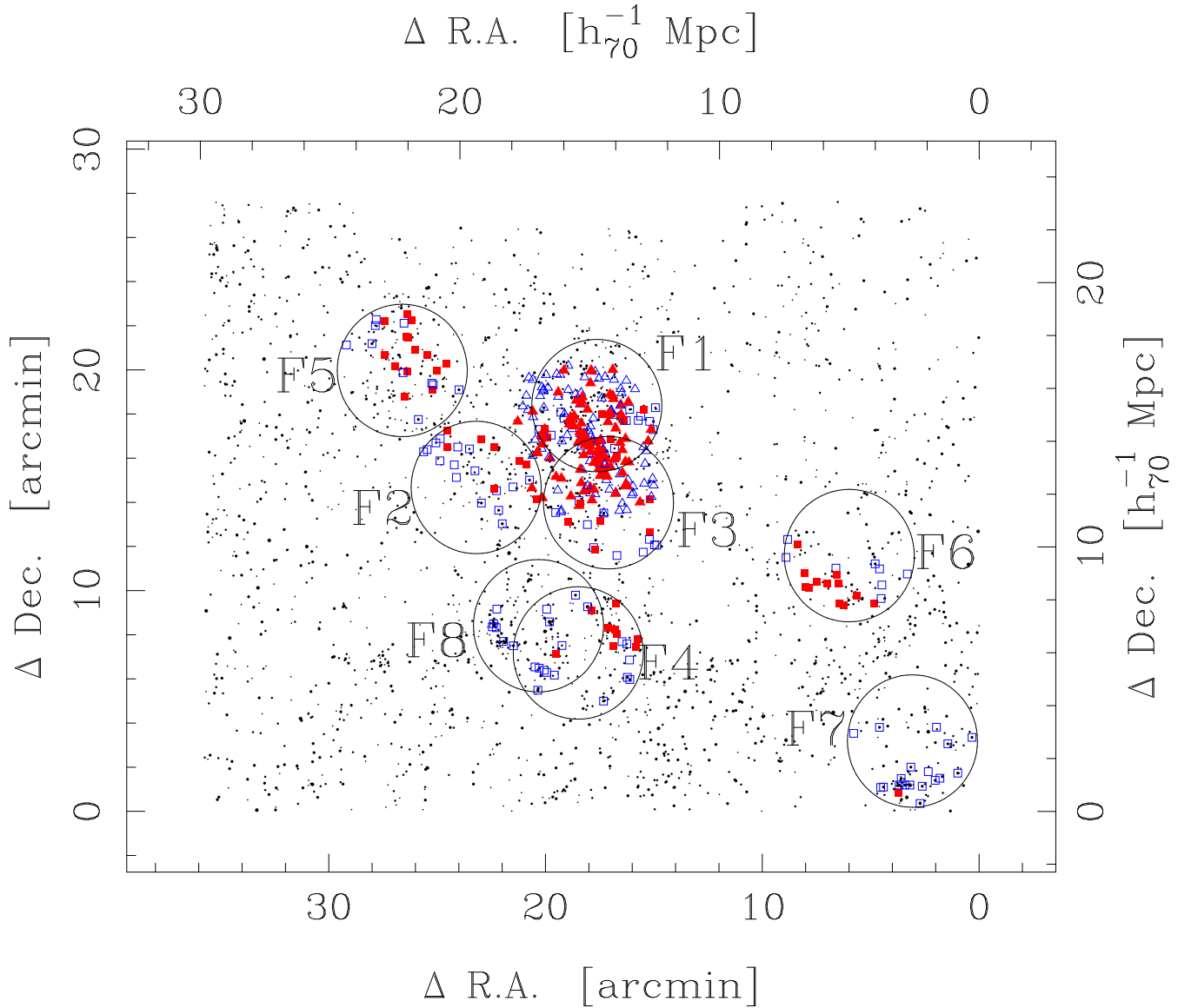}
\end{center}
\caption{
Spectroscopic target fields (F1-F8) as shown by the large circles.
The size of the large circles corresponds to the FoV of FOCAS.
The points show photo-$z$ selected galaxies at $0.76\leq z_{phot}\leq 0.88$.
The squares are our spectroscopic objects with secure redshift estimates ($z_{conf}=0$).
The triangles are spectroscopic objects drawn from \citet{demarco05} and \citet{jorgensen05}.
The filled squares and triangles show galaxies at $0.82<z_{spec}<0.85$, and the open
squares and triangles show galaxies outside of this redshift range.
The top and right ticks show the comoving scales.
}
\label{fig:target_fields}
\end{figure*}

We selected 8 FOCAS fields which efficiently cover the large-scale
structure traced by photometry as shown in Fig. \ref{fig:target_fields}.
Target galaxies were primarily selected on the basis of photometric redshifts.
Galaxies at $0.76\leq z_{phot}\leq 0.88$ were given the highest priority
in the mask design.
This broad and asymmetric redshift range was adopted so that we
do not miss the true cluster members whose photometric redshifts
are slightly wrong.
Roughly 60 percent of the targeted galaxies fall in this photo-$z$ range.
Each mask had about 40 slits.
Target galaxies are located towards one side of each FoV of FOCAS.
This is to optimize the wavelength coverage of a spectrum so that we can
observe the most important spectral features of galaxies 
at the cluster redshift (e.g. [OII] and CaII H+K).

\begin{figure}
\begin{center}
\leavevmode
\epsfxsize 1.0\hsize \epsfbox{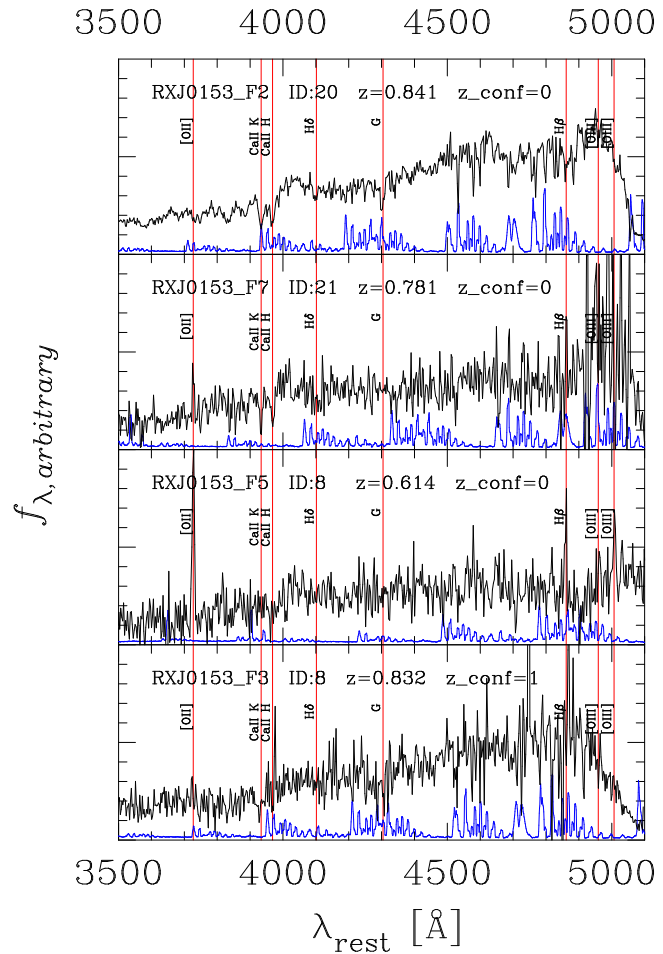}
\end{center}
\caption{
Examples of our spectra.  The sky emission lines are drawn at the bottom in each panel.
}
\label{fig:spec_example}
\end{figure}

Observing conditions were variable : moderate to poor seeing
and non-photometric conditions for some nights.
Total on-source exposures are listed in Table \ref{tab:obs_summary}.
Data reduction is performed in a standard manner using {\sc IRAF}.
We correct for the A-band telluric extinction by creating an absorption profile
from a composite spectra of high-$S/N$ galaxies.
We do not correct for the Galactic extinction.
All the reduced 1d/2d spectra are visually inspected using purpose written software.
Rough redshift estimates and redshift confidence flags ($z_{conf}$) are
then assigned to each object.
Objects with secure/probable redshift estimates are flagged as $z_{conf}=0$,
and objects with likely redshifts are flagged as $z_{conf}=1$.
The flag $z_{conf}=1$ is basically assigned to low-$S/N$ galaxies.
Important spectroscopic features (e.g. CaII H+K) are sometimes
heavily contaminated by the night sky emissions and in this case
we assigned $z_{conf}=1$.
We present in Fig. \ref{fig:spec_example} examples of our spectra.

Based on the rough redshift estimates, we fit a Gaussian profile
to emission/absorption features and redshifts are accurately measured.
The Gaussian fitting is repeated 100,000 times by resampling pixels around
the features and randomly adding noise, which is estimated directly from
each spectrum.
A 68 percentile interval of the redshift distribution is quoted as an error.
We note that this error should be considered as a rough estimate.
Note as well that it does not include an error in the wavelength calibration,
which is typically $\sim0.3\rm\AA$.
We obtain 161 secure redshifts ($z_{conf}=0$) out of 310 observed galaxies.
We present our spectroscopic catalogue in Table \ref{tab:spec_data}.
Astrometry is performed using the USNO2-B catalogue \citep{monet03}.
An accuracy of our coordinates is roughly $\sim0''.2$.

We cross-match our spectroscopic objects with those from \citet{demarco05}.
Six objects are matched.
Redshifts of 5 objects out of 6 are matched within errors, and 
the median of $z_{spec,demarco}-z_{spec,tanaka}$ is $-0.00004$
and the dispersion around it is 0.00068.
No systematic difference is seen.
However, one object has a deviant redshift:
$z_{spec,demarco}=0.8230$, while $z_{spec,tanaka}=0.7612$.
This object (Field ID: F1, Slit ID:9) is flagged $z_{conf}=1$ in our catalogue,
and hence our measurement may be wrong
(photo-$z$ is consistent with the Demarco et al.'s measurement).
In what follows, we do not use $z_{conf}=1$ galaxies.

We combine our spectroscopic catalogue with those from \citet{demarco05} and \citet{jorgensen05}.
This makes a large spectroscopic sample of 347 galaxies around the cluster.
Based on this large sample, we present large-scale structures surrounding the cluster
in the following section.

%--------------------------figures

%-------------

%---------------------------table
\begin{table}
\caption{
Log of the spectroscopic observation.
}
\label{tab:obs_summary}
\begin{tabular}{ll}
\hline
Field ID & Exposures\\
\hline
F1 & 1800s $\times$ 4 shots\\
F2 & 1800s $\times$ 4 shots $+$ 1345s $\times$ 1 shot\\
F3 & 1800s $\times$ 1 shot $+$ 1540s $\times$ 1 shot\\
F4 & 1800s $\times$ 4 shots\\
F5 & 1800s $\times$ 3 shots\\
F6 & 1800s $\times$ 4 shots\\
F7 & 1800s $\times$ 3 shots\\
F8 & 1800s $\times$ 2 shots\\
\hline
\end{tabular}
\end{table}

%---------------------------table
\begin{table*}
\caption{
Catalogue of our spectroscopic objects.
{\it This table will appear in its entirety in the electric edition of the journal.}
}
\label{tab:spec_data}
\begin{tabular}{llllrrrrllllll}
\hline
ID & R.A. & Dec. & $m_{z', tot}$ & $V-R$ & $R-i'$ & $i'-z'$ &
$z_{phot}$ & $z_{spec}$ & $z_{spec, min}$ & $z_{spec, max}$ & $z_{conf}$\\
\hline
F1-3 & 1 52 51.66 & -13 57 12.35 & 24.02 & 0.30 & 0.20 & 0.22 & 1.52 & 0.3150 & 0.3150 & 0.3151 &  0\\
F1-4a & 1 52 51.14 & -13 57 29.78 & 22.75 & 1.22 & -2.06 & 0.22 & 0.00 & 0.5165 & 0.5165 & 0.5166 &  0\\
F1-4b & 1 52 50.96 & -13 57 30.10 & 21.78 & 1.03 & 1.01 & 0.72 & 0.88 & 0.8455 & 0.8448 & 0.8462 &  0\\
F1-5 & 1 52 50.51 & -13 57 12.09 & 21.87 & 1.15 & 1.00 & 0.66 & 0.85 & 0.8350 & 0.8344 & 0.8354 &  0\\
F1-6 & 1 52 50.06 & -13 57 34.89 & 22.64 & 1.17 & 1.13 & 0.63 & 0.80 & 0.8406 & 0.8404 & 0.8408 &  0\\
F1-7 & 1 52 49.29 & -13 57 27.79 & 21.47 & 1.18 & 1.03 & 0.84 & 0.93 & 0.9588 & 0.9577 & 0.9601 &  0\\
F1-9 & 1 52 47.98 & -13 55 31.11 & 22.17 & 0.73 & 0.89 & 0.49 & 0.83 & 0.7612 & 0.7606 & 0.7616 &  1\\
F1-10 & 1 52 47.45 & -13 56 25.81 & 22.85 & 0.26 & 0.47 & 0.37 & 1.11 & 1.0540 & 1.0540 & 1.0541 &  0\\
F1-11 & 1 52 46.86 & -13 56 42.64 & 23.49 & 0.26 & 0.59 & 0.39 & 1.02 & 1.0552 & 1.0551 & 1.0553 &  0\\
F1-14 & 1 52 45.35 & -13 57 7.94 & 23.69 & 0.86 & -0.06 & -0.09 & 0.36 & 3.9280 & 3.9279 & 3.9280 &  0\\
F1-$15^*$ & 1 52 45.18 & -13 57 4.26 & 23.59 & 0.96 & 0.09 & 0.19 & -- & 3.9276 & 3.9275 & 3.9277 &  0\\
F1-16 & 1 52 44.45 & -13 56 56.02 & 23.95 & 0.90 & -0.08 & -0.05 & 0.38 & 3.9277 & 3.9277 & 3.9278 &  0\\
F1-21 & 1 52 41.13 & -13 57 43.03 & 21.31 & 1.29 & 1.20 & 0.79 & 0.87 & 0.8323 & 0.8318 & 0.8328 &  0\\
\hline
\end{tabular}
\begin{flushleft}
The format of ID is ``Field ID -- Slit ID''.
Total magnitudes ($m_{z',tot}$) are measured in Kron-type apertures ({\sc MAG\_AUTO}),
while colours are measured within $2''$ apertures.
Objects marked $^*$ are missing photometric redshifts due to strong blending with nearby objects.
\end{flushleft}
\end{table*}

%
%
%------------------------------------------------------------LARGE-SCALE STRUCTURES
\section{Large-Scale Structures}
\label{sec:lss}

We first present distribution of spectroscopically observed galaxies.
We then point to the question how accurate our photometric redshifts are.
In particular, we assess the colour dependence of the accuracy of photometric redshifts.

%-------------------------lss at z=0.8
\subsection{Large-Scale Structures at $z\sim0.8$}
\label{sec:lss_z08}

\begin{figure*}
\begin{center}
\leavevmode
\epsfxsize 0.44\hsize \epsfbox{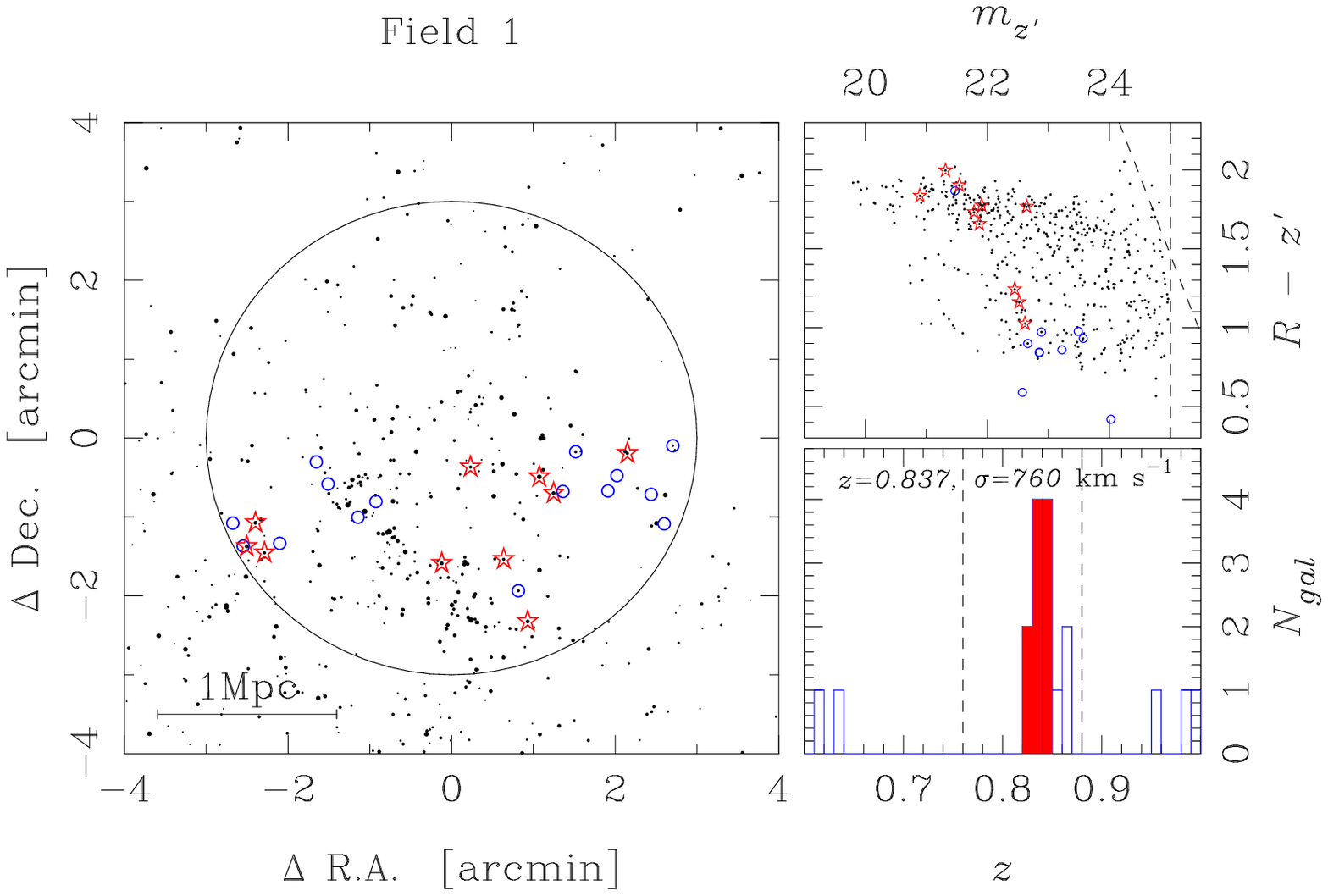}\hspace{0.5cm}
\epsfxsize 0.44\hsize \epsfbox{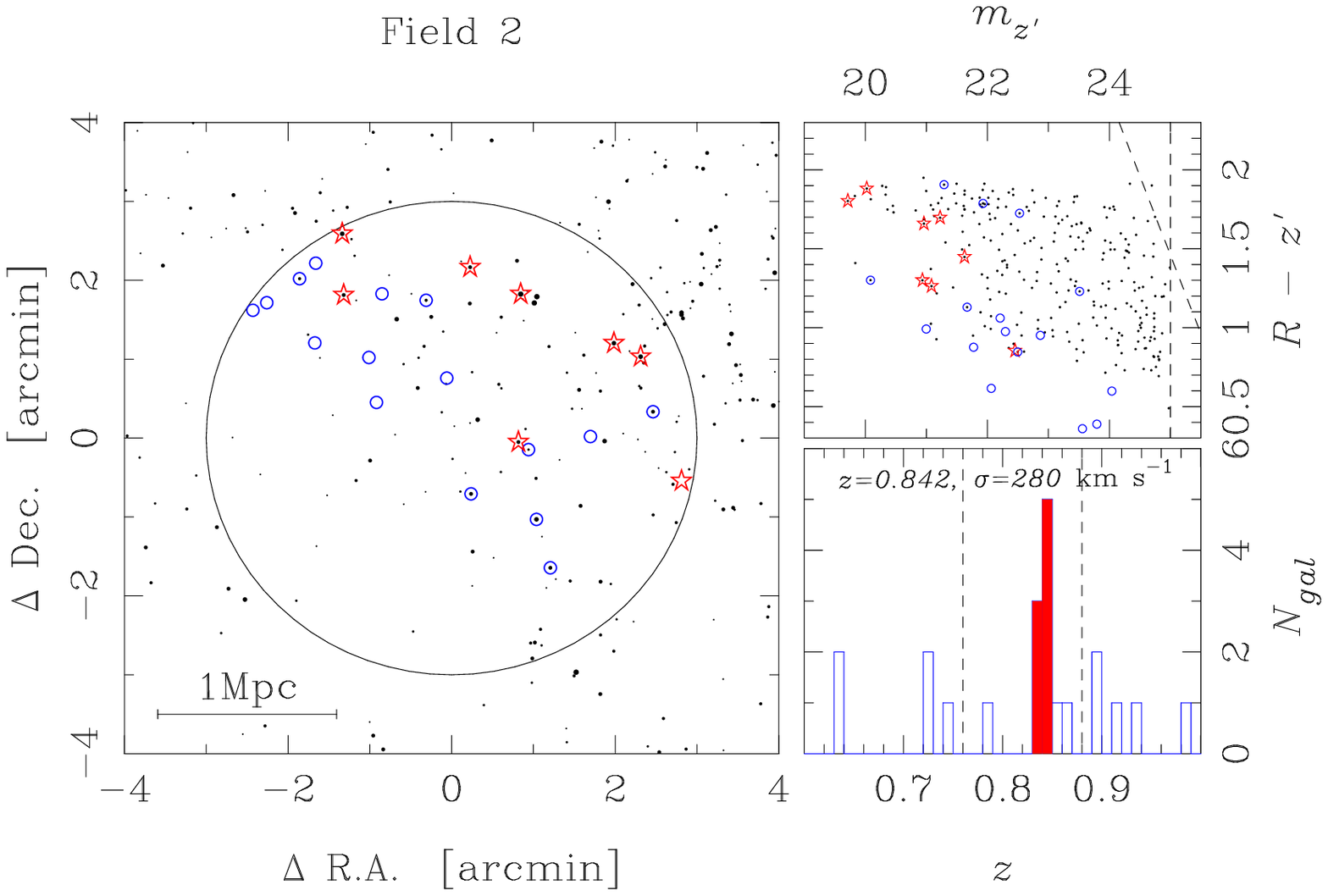}\\\vspace{0.1cm}
\epsfxsize 0.44\hsize \epsfbox{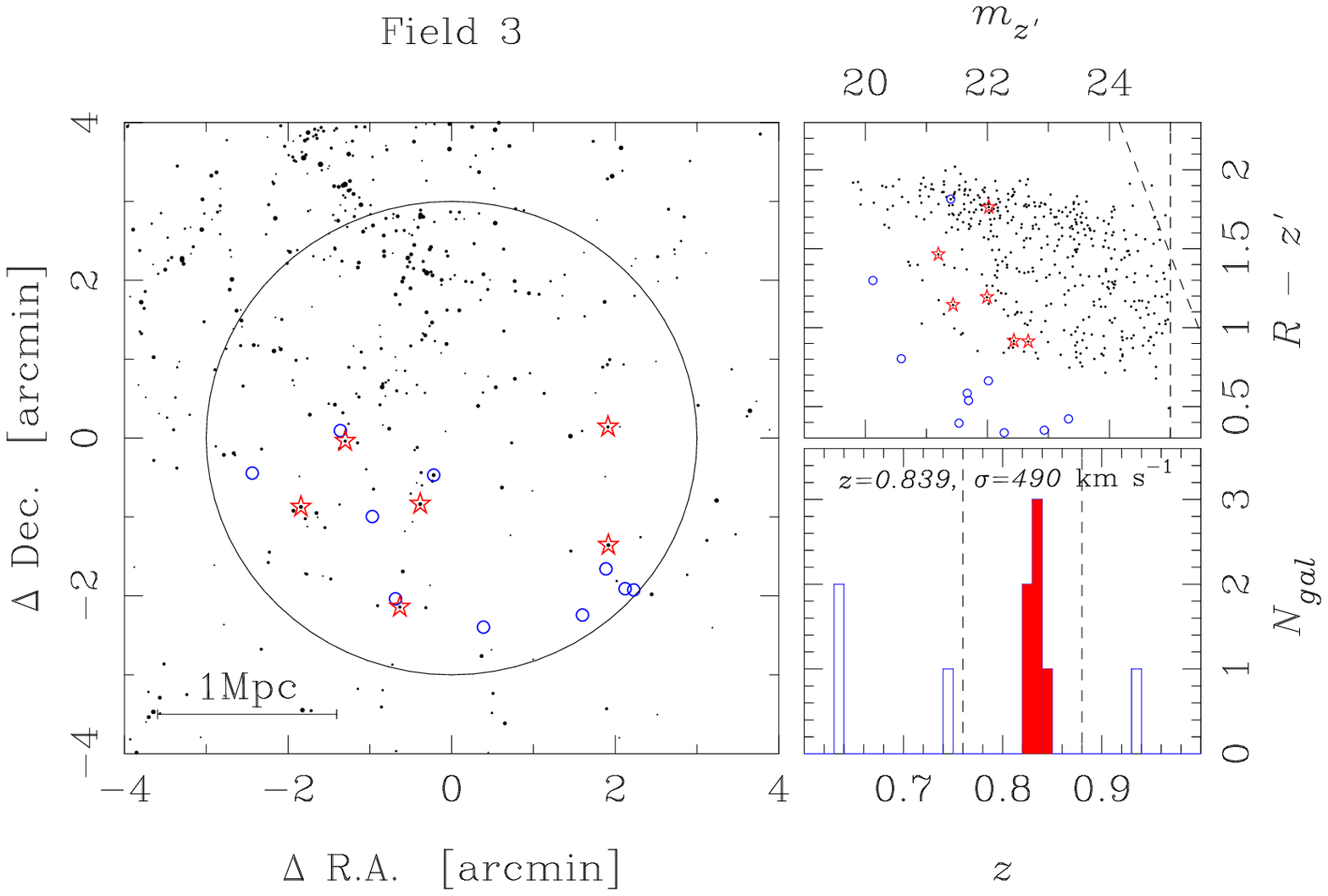}\hspace{0.5cm}
\epsfxsize 0.44\hsize \epsfbox{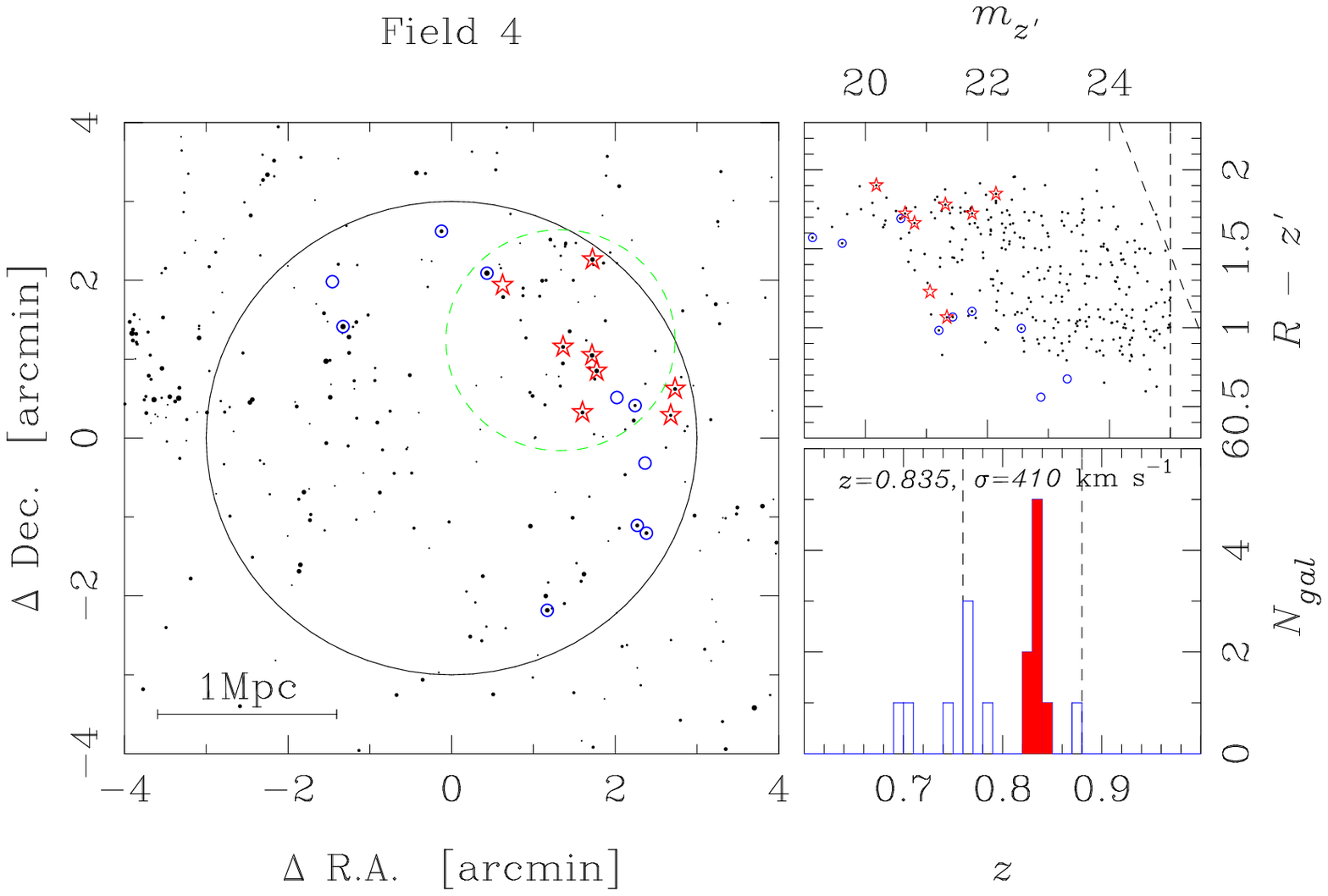}\\\vspace{0.1cm}
\epsfxsize 0.44\hsize \epsfbox{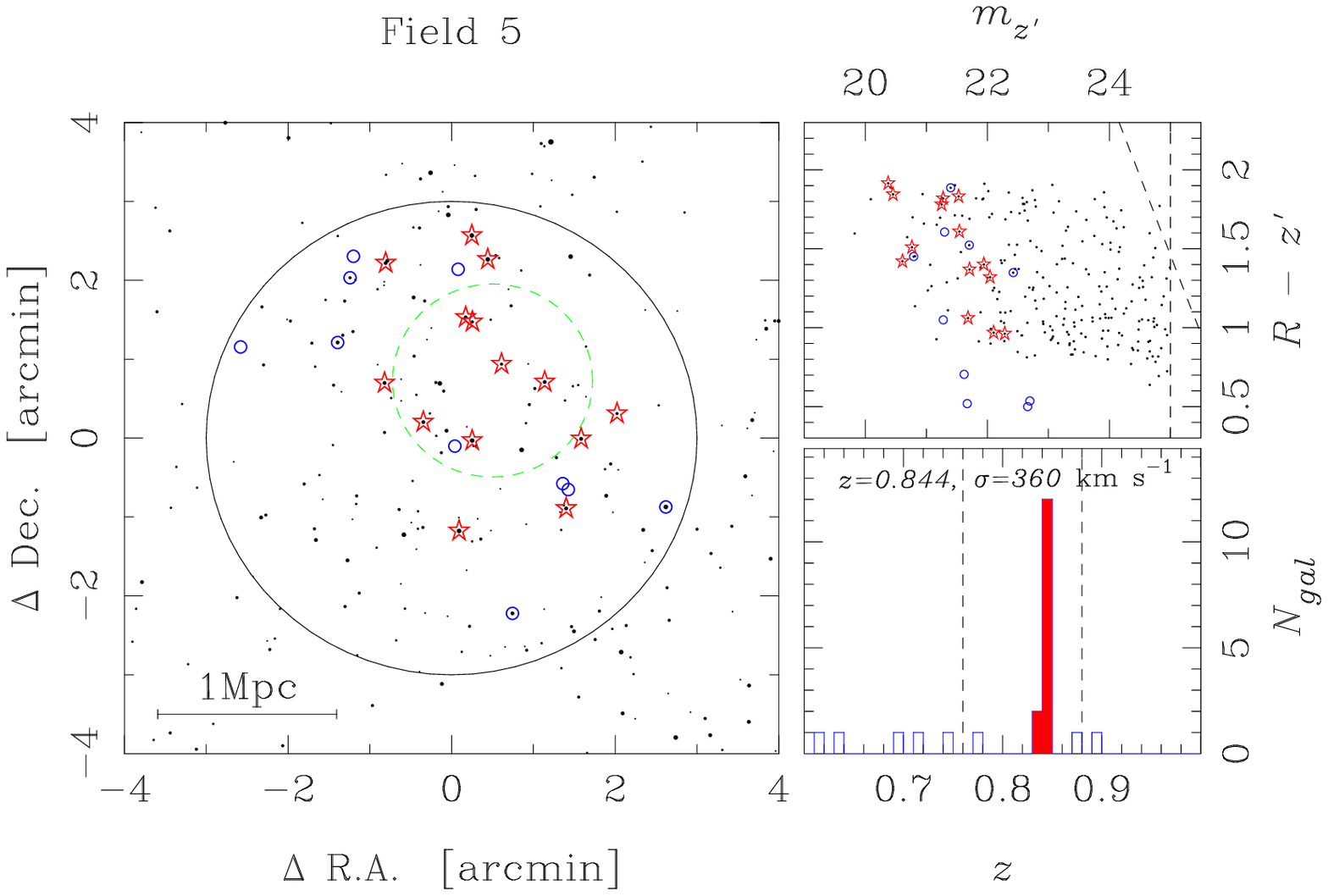}\hspace{0.5cm}
\epsfxsize 0.44\hsize \epsfbox{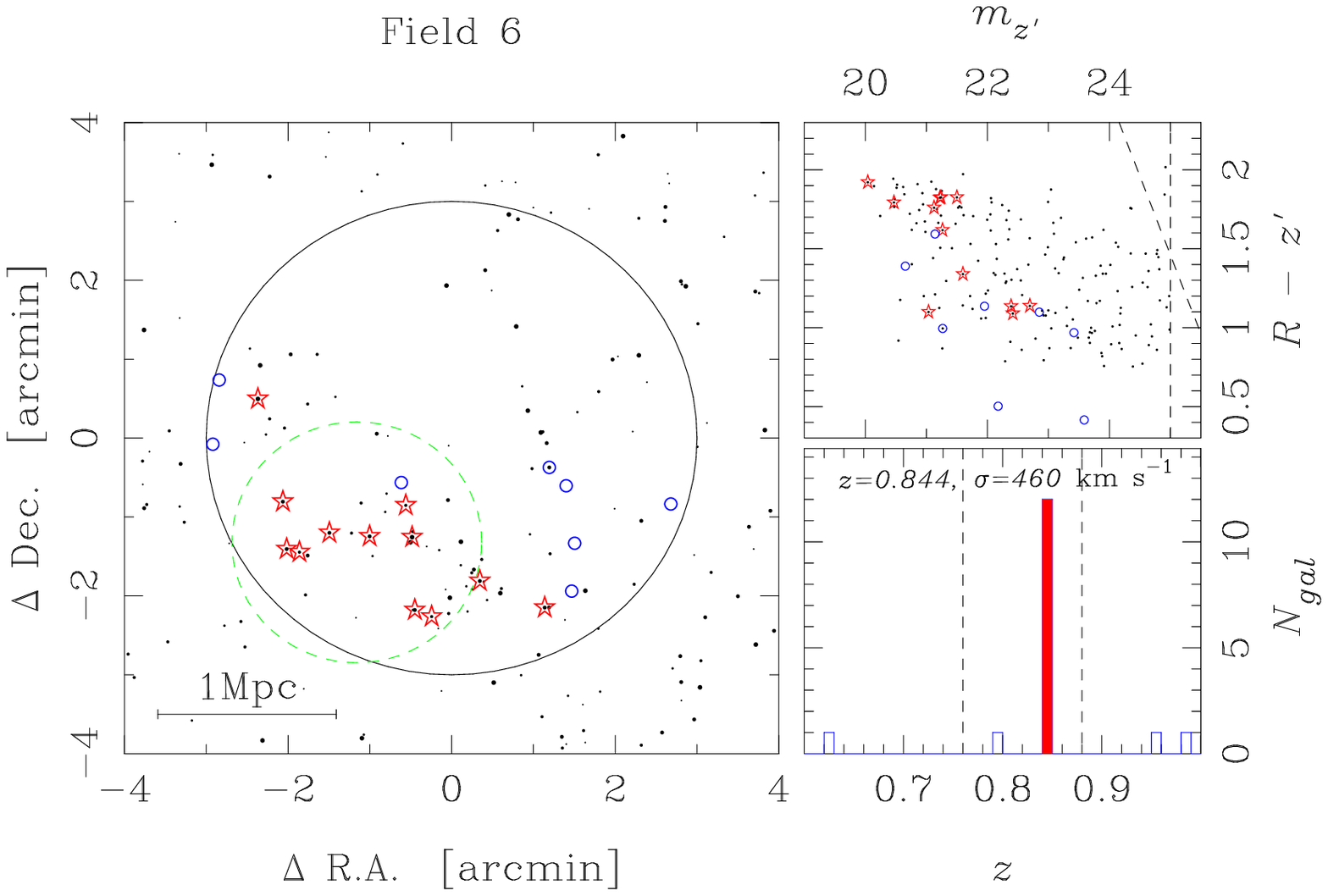}\\\vspace{0.1cm}
\epsfxsize 0.44\hsize \epsfbox{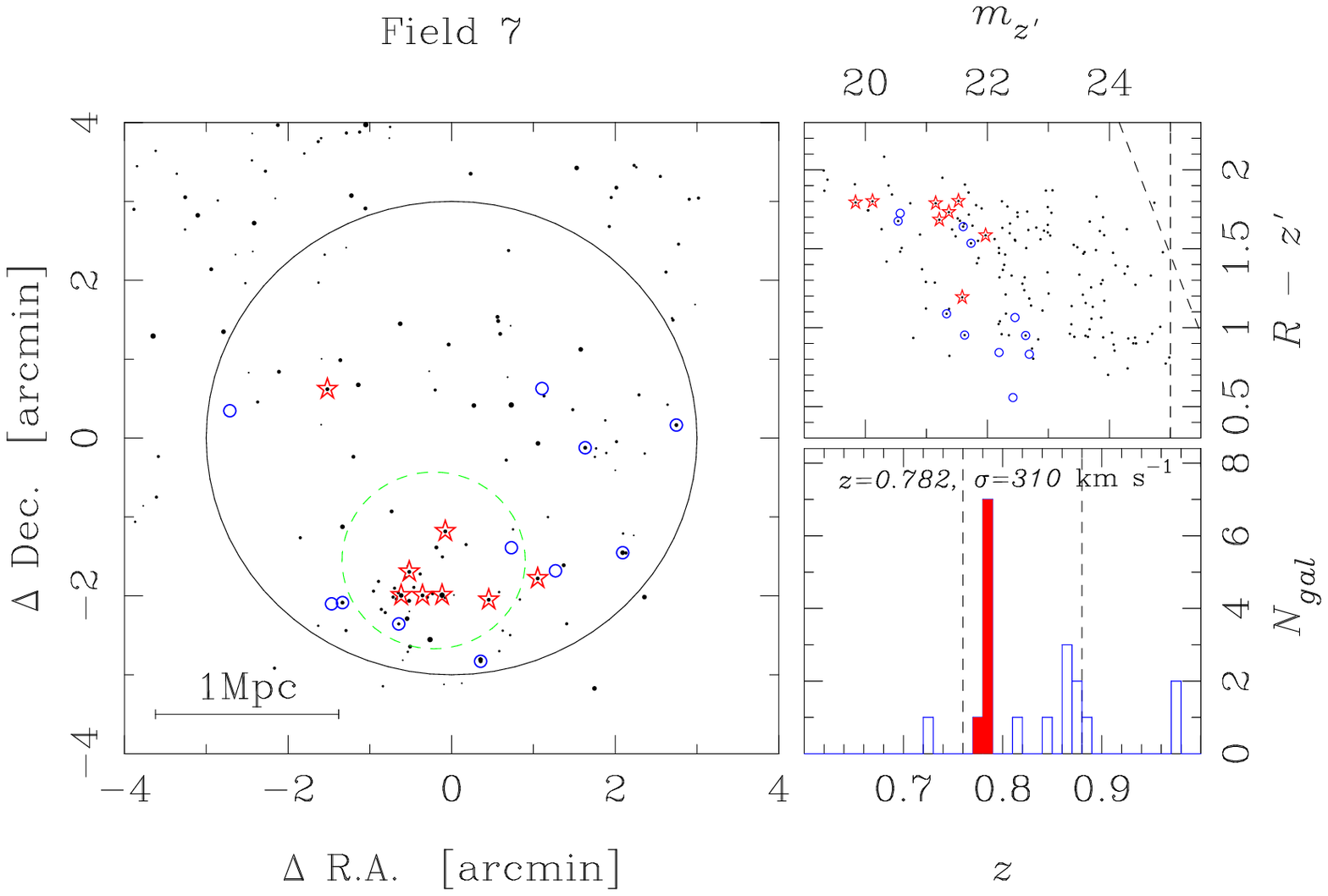}\hspace{0.5cm}
\epsfxsize 0.44\hsize \epsfbox{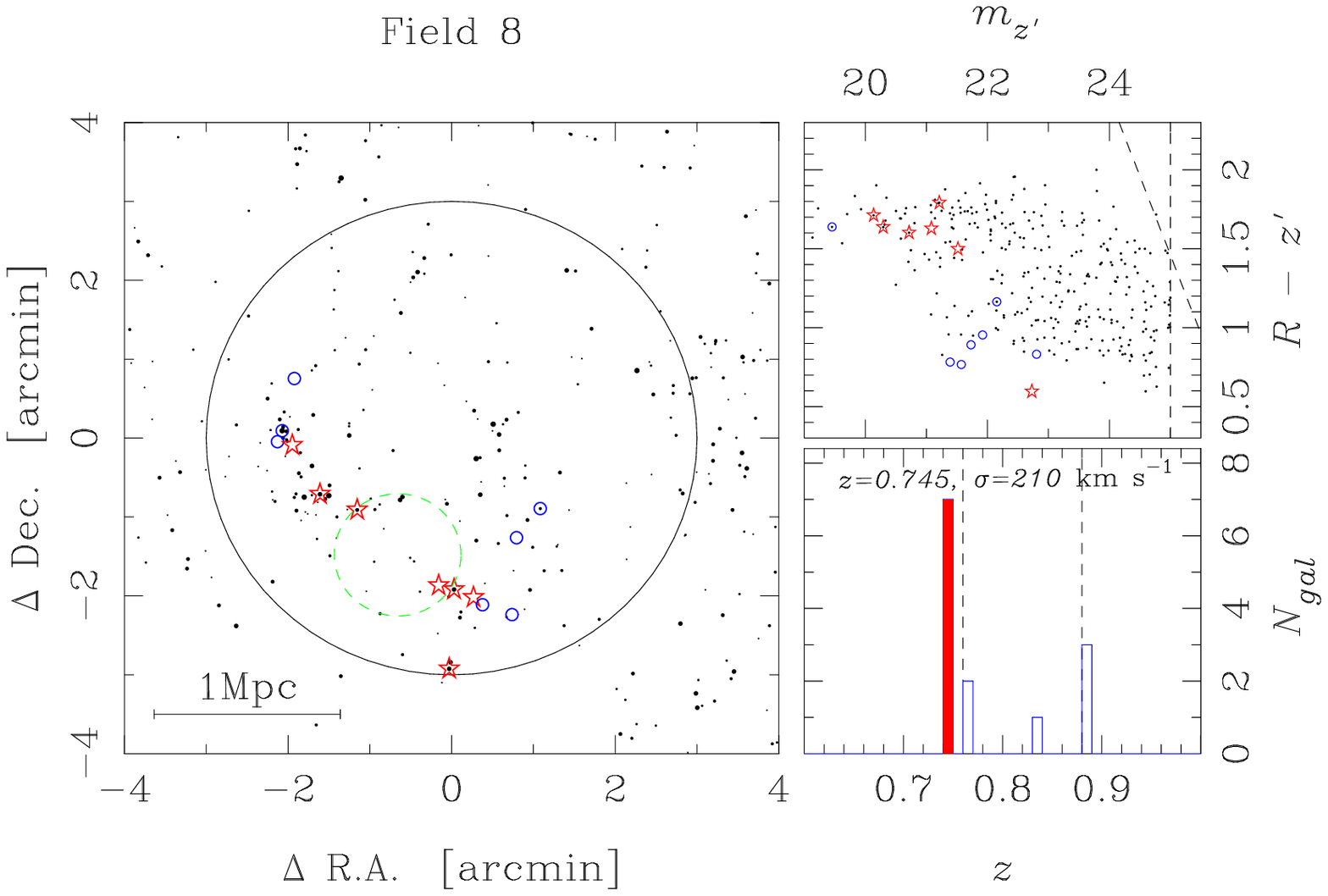}\\
\end{center}
\caption{
Close-up views of the target fields.
{\it Bottom-right panel in each plot:}
The redshift distribution of spectroscopically observed galaxies in each field.
The filled histograms show a redshift spike and galaxies in this spike
are shown as stars in other panels.
The vertical dashed lines mean our primary photo-$z$ selection range.
{\it Top-right  panel in each plot:} The $R-z'$ colour plotted against the $z'$-band magnitude
using galaxies at $0.76\leq z_{phot}\leq0.88$.  The stars are spectroscopic objects in
the redshift spike, and the open circles are objects outside of this spike.
The dashed lines show $5\sigma$ limiting magnitudes and colours.
{\it Left panel in each plot:} The points show the distribution of photo-$z$ selected
($0.76\leq z_{phot}\leq0.88$) galaxies, and the circle shows the FoV of FOCAS.
The 1 Mpc scale is expressed as the physical distance.
The stars/circles are galaxies inside/outside the redshift spike, respectively.
The virial radius of a group/cluster is plotted as the dashed circle in some plots.
}
\label{fig:close_up_views}
\end{figure*}

Fig. \ref{fig:close_up_views} shows our results in each field.
Note that only our spectroscopic objects are plotted in Fig.
\ref{fig:close_up_views}, and the data from \citet{demarco05} and
\citet{jorgensen05} are not used.
It is impressive that the redshift distribution of spectroscopically observed
galaxies show a sharp redshift spike in all the fields.
Although galaxies are selected primarily by $z_{phot}$ in the mask design,
the photo-$z$ selection range ($0.76\leq z_{phot}\leq0.88$) is much wider than
the observed redshift spikes.
Therefore, the redshift spikes are not a product of selection effects
but are real structures in each field.
Furthermore, the peaks are located at or near the cluster redshift
of $z=0.837$ \citep{demarco05}, except for F7 and F8
(these turn out to be foreground systems).
This is strong evidence for large-scale structures at $z\sim0.83$.
Table \ref{tab:vel_disp} presents the redshift centres of the redshift spikes.
Fig. \ref{fig:target_fields} shows the distribution of galaxies at $0.82<z_{spec}<0.85$.
The spectroscopic samples from \citet{demarco05} and \citet{jorgensen05}
are included in this plot.
As shown by the triangles, \citet{demarco05} and \citet{jorgensen05} focused only
on the central part of the cluster,
while we mainly explore the surrounding regions avoiding duplication of
the spectroscopic targets.
We note that 154 galaxies out of the 347 galaxies with secure redshifts
lie at $0.82<z_{spec}<0.85$.

The redshift peaks in F1, F3, and F4, are located
very close to the redshift of the main cluster ($z=0.837$), and
forming a large filament over 12~Mpc (comoving) in the N--S direction.
Interestingly, there seems to be another large scale filament at $z\sim0.844$
in the direction of NE--SSW traced by F5, F2, and F6.
This second filament includes a subclump at $z=0.845$
located at $\sim3$ arcmin eastward of the main cluster discovered by
the previous studies \citep{demarco05,jorgensen05,girardi05}.
It turns out that this second filament also extends to more than
20~Mpc (comoving).
Therefore, it seems that there are two large sheets/filaments of galaxies
crossing in this field.
The relative velocity in the line of sight of these two sheets is
1000 km s$^{-1}$, and it is likely that they are physically connected.
It is interesting to recall that the main cluster itself
is in the process of mergers \citep{maughan03,demarco05,jee05}.
This cluster is therefore currently experiencing the vigorous assembly of the
surrounding systems along the filaments and will experience more mergers
in the near future.
Unfortunately, however, the large errors in the velocity dispersion
estimates (i.e. mass estimates) with the current data set
do not allow us to perform detailed dynamical analysis.

Based on $z_{phot}$, \citet{tanaka05} identified galaxy groups in F4, F5, F6, and F7,
and a galaxy cluster in F8.  We examine these fields in detail.
We measure velocity dispersions of the redshift spikes using the
gapper method \citep{beers90}.
The results are shown in Table \ref{tab:vel_disp} and the bottom-right corner of 
each panel in Fig. \ref{fig:close_up_views}.
We confirm the biweight estimator gives consistent estimates \citep{beers90}.
Errors are estimated from the jackknife resampling of the spectroscopic members.
From these velocity dispersions, we evaluate virial radii \citep{carlberg97}
as plotted in the figure.
Here we adopt $r_{200}$ as a virial radius
($r_{200}$ is the radius within which the mean interior density is
200 times the critical density of the Universe).
The spatial centre of groups/clusters are determined from the distribution of
spectroscopic galaxies weighted by an inverse of redshift difference from
the central redshift of the groups/clusters.

Galaxies (including photo-$z$ selected galaxies) in F4, F6, and F7
show clear spatial clustering (see Fig. \ref{fig:target_fields} as well)
and most of the galaxies are distributed within the virial radius
of each group.
This suggests that these are gravitationally bound systems.
On the other hand, the spatial clustering of galaxies in F5 is less strong.
This may not be a bound system.
In F8, galaxies are clearly clustered.  However, the estimated virial radius is
too small for its apparent extent.  This is probably because the spectroscopic
sampling of galaxies in this field is too sparse for us to make an accurate
measure of their velocity dispersion.
In fact, a number of photometric members in this system
is the largest among others, excluding the main cluster at $z=0.83$.
We regard systems in F4, F6, and F7 as groups and the one in F8 as a cluster.
We do not include F5 in groups since it is not clear if this system is
gravitationally bound or not.
The fact that F5 is dominated by blue galaxies may suggest that this is not a bound system.
But, we have confirmed that the inclusion of F5 does not alter our conclusions.
We recall that this classification of group and cluster environments is
based on 2-dimensional galaxy density on both local and global scales \citep{tanaka05}.
We cannot define groups and clusters based on the spectroscopic data in hand
(e.g. velocity dispersions) due to the poor statistics.
Star formation histories of galaxies in these regions are examined in section \ref{sec:sfh}.

We note that we serendipitously discover three Ly$\alpha$ emitters at $z=3.928$ in F1.
These galaxies show very similar broadband colours and lie at the same redshift,
and it turns out that these are originally a single galaxy strongly lensed by
the main cluster at $z=0.83$.
Details of the galaxies and the lensing models are reported in \citet{umetsu05}.

%--------------------------figures

%--------------------------table
\begin{table*}
\caption{
Redshifts and velocity dispersions of the redshift spike in each field.
$N_{member}$ is a number of spectroscopic galaxies
in the redshift spike.
The central redshifts and velocity dispersions are measured using the biweight estimator
and the gapper method, respectively.
Errors are estimated from the jackknife resampling.
}
\label{tab:vel_disp}
\begin{tabular}{ccccl}
\hline
Field ID & $N_{member}$ & $z$ & $\sigma\rm\ [km\ s^{-1}]$ & environment\\
\hline
F1 & 10 & $0.8368\pm0.0027$ & $756\pm145$  & main cluster\\
F2 & 8  & $0.8416\pm0.0013$ & $277\pm30 $  & filament\\
F3 & 6  & $0.8385\pm0.0047$ & $490\pm449$  & filament\\
F4 & 8  & $0.8348\pm0.0009$ & $413\pm241$  & group\\
F5 & 14 & $0.8439\pm0.0017$ & $362\pm159$ & group or filament\\
F6 & 12 & $0.8443\pm0.0046$ & $457\pm46 $  & group\\
F7 & 8  & $0.7823\pm0.0012$ & $313\pm113$  & foreground group\\
F8 & 7  & $0.7453\pm0.0006$ & $210\pm98 $  & foreground cluster\\
\hline
\end{tabular}
\end{table*}

%--------------------------photo-z accuracy
\subsection{Accuracy of Photometric Redshifts}
\label{sec:photoz}

\begin{figure}
\begin{center}
\leavevmode
\epsfxsize 1.0\hsize \epsfbox{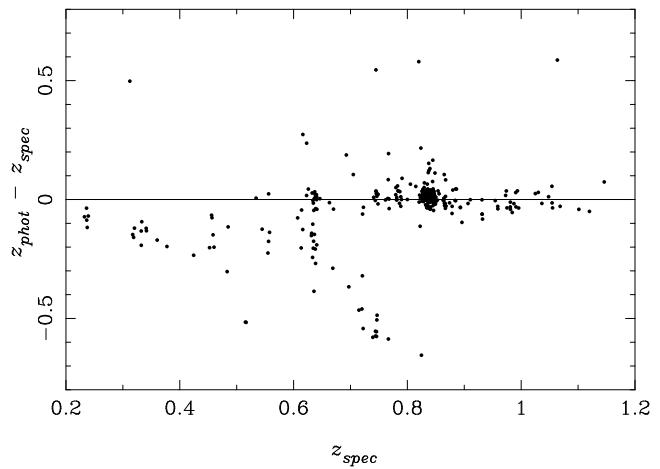}
\end{center}
\caption{
Differences between $z_{phot}$ and $z_{spec}$ plotted against $z_{spec}$.
}
\label{fig:photoz_accuracy}
\end{figure}

With the 347 spectroscopic redshifts now available in this field,
we assess the accuracy of our photometric redshifts.
The errors in photometric redshift ($\Delta z =z_{phot}-z_{spec}$)
is plotted against $z_{spec}$ in Fig. \ref{fig:photoz_accuracy}.
Although our photometric redshifts are relatively accurate especially
around the cluster redshift, some clear deviations can be seen.
In particular, some galaxies at $0.7\lesssim z_{spec}\lesssim0.8$
significantly deviate ($\Delta z\sim-0.6$).
Also, galaxies at $z\sim0.6$ show a lower redshift tail in their $z_{phot}$
distribution.
At $z_{spec}\lesssim0.6$, there is a systematic offset between
$z_{spec}$ and $z_{phot}$.

\begin{figure*}
\begin{center}
\leavevmode
\epsfxsize 0.45\hsize \epsfbox{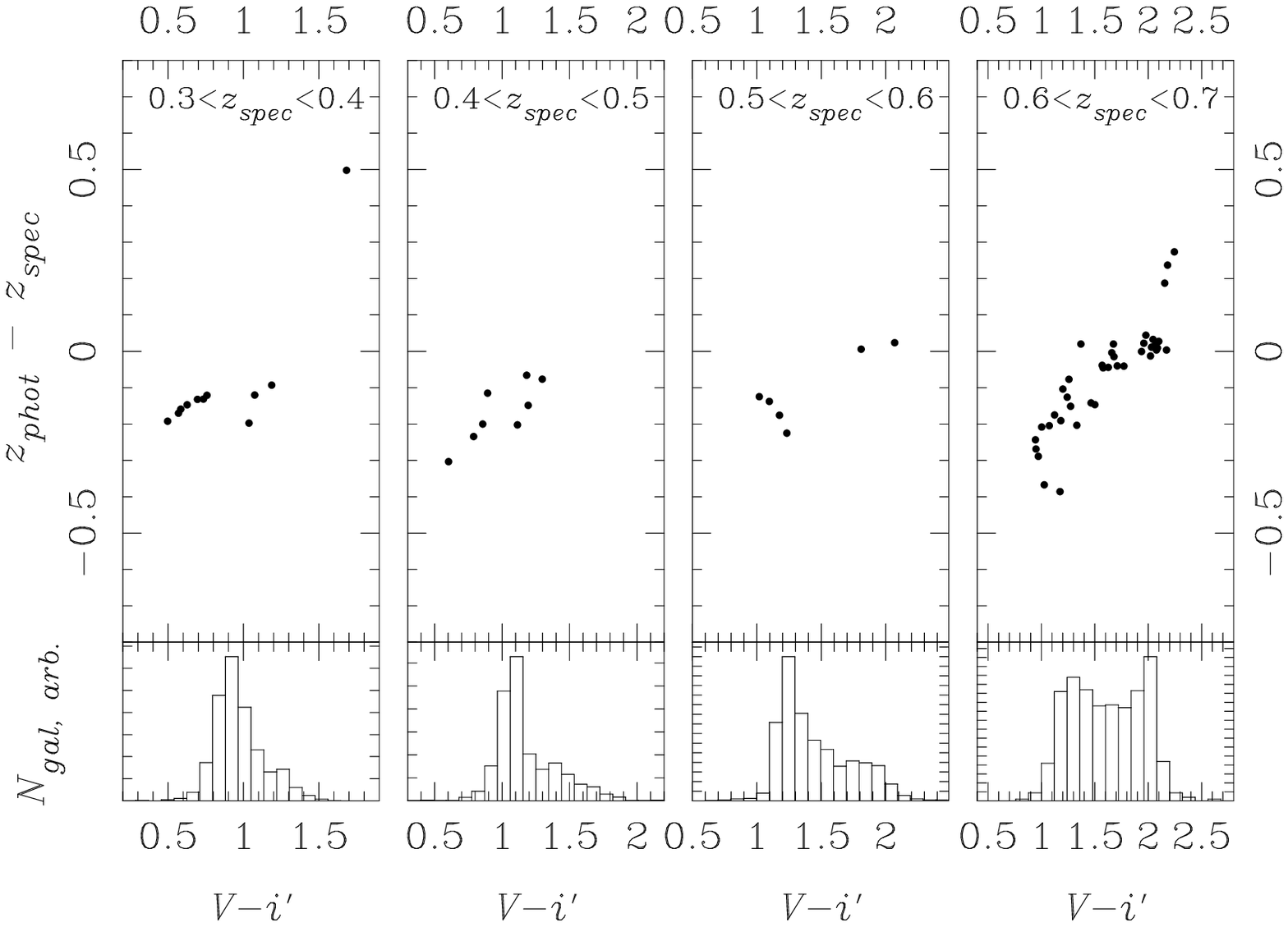}\hspace{0.5cm}
\epsfxsize 0.45\hsize \epsfbox{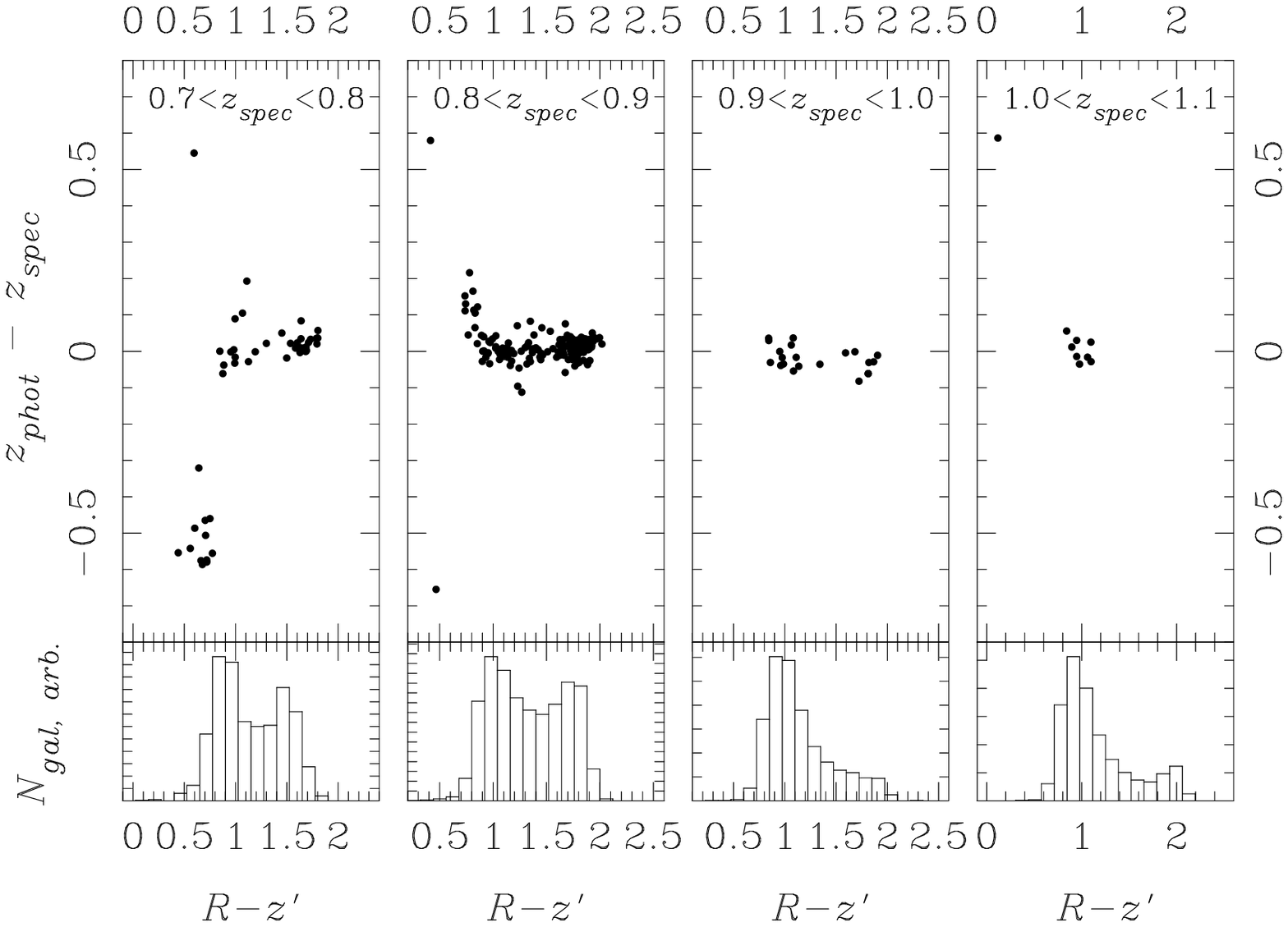}
\end{center}
\caption{
The accuracy of photometric redshifts as a function of galaxy colours.
The sample is divided into eight redshift bins based on the spectroscopic redshifts.
At each redshift bin, the upper panel shows $z_{phot}-z_{spec}$
against galaxy colours.
The bottom panel presents the colour distribution of photo-$z$ selected galaxies.
}
\label{fig:photoz_colour_dep}
\end{figure*}

We find that the deviations are related to intrinsic SEDs of galaxies.
Fig. \ref{fig:photoz_colour_dep} shows $\Delta z$ as a function of
galaxy colours.
The figure demonstrates that $z_{phot}$ is fairly accurate for red galaxies
at all redshifts under study, while it is less accurate for blue galaxies.
At $0.8<z<1.1$, only the bluest galaxies strongly deviate, but at $0.6<z<0.7$,
blue-intermediate colour galaxies also deviate.
In Fig. \ref{fig:photoz_accuracy}, we find that the
galaxies at $z_{spec}\lesssim0.6$ have systematically lower $z_{phot}$.
It turns out that most of these galaxies are blue galaxies as shown
in Fig. \ref{fig:photoz_colour_dep}.
A few very red galaxies show deviant redshifts
(e.g. three red galaxies at $0.6<z_{spec}<0.7$).
We find that these galaxies show slightly redder colours
than our reddest model colours.
These are likely to be heavily dust obscured galaxies,
though such population is rare.

The deviation of the bluest galaxies can be explained by the fact that
we lack model templates for such very blue galaxies \citep{kodama99}.
The bluest galaxies cannot be fit by any of the templates and
thus the photo-$z$ outputs very deviant redshifts.
As for the deviation of the galaxies with blue-to-intermediate colours
at $z\lesssim0.7$, we speculate that our filter set ($VRi'z'$) is no longer
optimal.
We do not cover a wide enough wavelength range (particularly in UV)
to reliably estimate redshifts of blue galaxies that do not show prominent
SED features \citep{kodama99}.  Adding $B$-band will improve the situation.

From the results presented in our previous paper
(see Fig. 1 of \citealt{tanaka05}),
we have expected that only the galaxies with intermediate colours would
show deviant photometric redshifts at the cluster redshift.
We do not, however, observe this trend in our large spectroscopic sample.
At the cluster redshift, even the intermediate colour galaxies
are given reasonably accurate photometric redshifts, and only the bluest
galaxies tend to deviate.
It may be the case that the deviation of intermediate colour galaxies could
be characteristic of $z\sim0.5$ galaxies (for which we saw the trend in
\citealt{tanaka05}).
Given the small number of spectroscopic galaxies at that redshift
in our current sample, we cannot pursue this issue further at this point.

The colour dependence tells us that we have to be careful when discussing
a fraction of red/blue galaxies (e.g. Butcher-Oemler effect) based on
photo-$z$.
It is expected to be a general trend that the photometric redshifts are
less accurate for blue galaxies than for red galaxies due to the less
strong SED features.
This colour dependent photo-$z$ accuracy would tend to underestimate
a fraction of blue galaxies, and could weaken evolutionary trends.
We note, however, that this colour dependence does not change
the primary conclusions presented in \citet{tanaka05}.
They reported the build-up of the colour-magnitude relation
from $z=0.83$ to $z=0$ in field, group, and cluster environments.
Their results are not strongly affected since they focused on red galaxies
whose photometric redshifts are accurate.
Further discussions can be found in that paper.

\begin{figure}
\begin{center}
\leavevmode
\epsfxsize 1.0\hsize \epsfbox{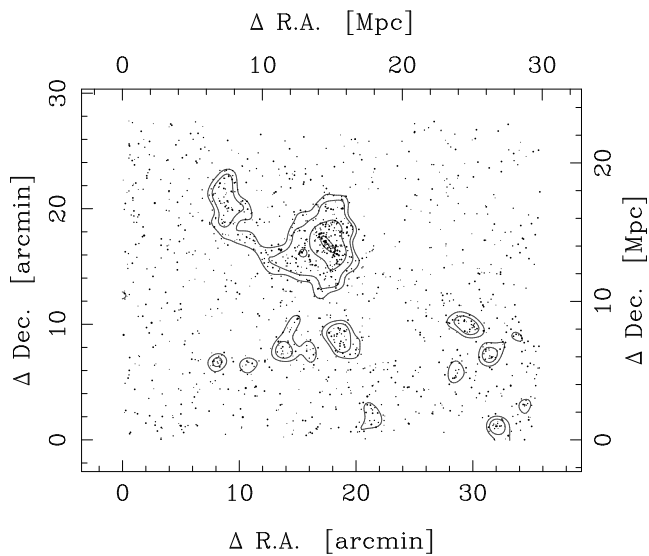}
\end{center}
\caption{
Distribution galaxies with $z_{cl}-0.03<z_{phot}<z_{cl}+0.05$ and $R-z'>1$.
The top and right ticks show the comoving scales.
The contours present 90, 95, 99, and 99.9\% of the density distribution in this field.
}
\label{fig:photoz_distrib}
\end{figure}

Finally, we revisit the distribution of photo-$z$ selected galaxies
with more optimal photo-$z$ and colour cuts.
Following the true error distribution shown in
Fig. \ref{fig:photoz_colour_dep},
we select galaxies with $z_{cl}-0.03<z_{phot}<z_{cl}+0.05$
($z_{cl}=0.837$; \citealt{demarco05}) and $R-z'>1$ and plot their
distribution in Fig. \ref{fig:photoz_distrib}.
The filamentary structure to the NE direction is more pronounced,
and it emerge from the SE of the main cluster and continuously extends
towards NE out to 10~Mpc.
Galaxy groups at South and SW of the main cluster are also confirmed.
They are likely to infall onto the cluster in the near future.
We stress that this is one of the most spectacular and reliable
large-scale structures of the $z\sim1$ Universe reaching down to
very faint magnitude ($M_V^*$+4)
traced by spectroscopically calibrated photometric redshifts.

%--------------------------figures

%-------------

%-------------

%
%
%------------------------------------------------------------STAR FORMATION HISTORIES
\section{Star Formation Histories}
\label{sec:sfh}

The concentrations of galaxies, clusters and groups, 
are spectroscopically identified in section \ref{sec:lss_z08}.
In the following, we infer star formation histories of galaxies from
their spectroscopic signatures as a function of environment.
Here we use our spectra only.
\citet{demarco05} and \citet{jorgensen05} focused only on the main cluster
and data for other environments (groups in particular) are not available.

%----------------------------
\subsection{Composite Spectra}
\label{sec:composite_spectra}
A typical $S/N$ of our spectra is not high enough to quantify the
spectroscopic properties of galaxies on an individual basis.
We thus composite a number of spectra to make a high $S/N$ spectrum
as was done by, for example, \citet{dressler04}.
Here we restrict ourselves to the red galaxies (those having
$|\Delta (R-z')|<0.2$ with respect to the colour-magnitude relation).
We do not examine blue galaxies because photometric redshifts are not
accurate for blue galaxies, and we cannot avoid a selection bias
(e.g. we miss very blue galaxies).
Each spectrum is normalized to unity at $\rm4000-4200\AA$ in the restframe,
where a spectrum is relatively flat.
A composite spectrum is then made by taking 2$\sigma$ clipped mean.
We combine the red galaxies in F1 and F8 to make a representative spectrum
of ``cluster'' galaxies.
A composite ``group'' spectrum is made from the red galaxies in F4, F6, and F7.
We also make a composite ``field'' spectrum from the isolated red galaxies
outside of the redshift spikes in each field but within $0.74<z_{spec}<0.88$.
The cluster/group/field spectra are constructed from 11/16/8 galaxies,
respectively.

Before comparing the composite spectra, we discuss the similarity 
in the photometric properties of the red galaxies among different
environments that are used to construct the cluster/group/field
composite spectra.
After correcting for the small k-correction and passive evolution effects
for galaxies at slightly different redshift \citep{kodama97},
we apply the 2-dimensional K-S test \citep{fasano87} for the distribution
of galaxies on the colour-magnitude diagram.
It does not reject the hypothesis that the colours and magnitudes of the
``cluster'' red galaxies and those of the ``group'' red galaxies are drawn
from the same parent population ($80\%$).
Therefore, the red galaxies in clusters and those in groups share
the common photometric properties and the only obvious difference is their
environment.
However, the K-S probability is found to be smaller between
``cluster'' and ``field'' ($20\%$) and ``group'' and ``field'' ($20\%$).
Hence, field red galaxies may have slightly different photometric properties.

\begin{figure}
\begin{center}
\leavevmode
\epsfxsize 1.0\hsize \epsfbox{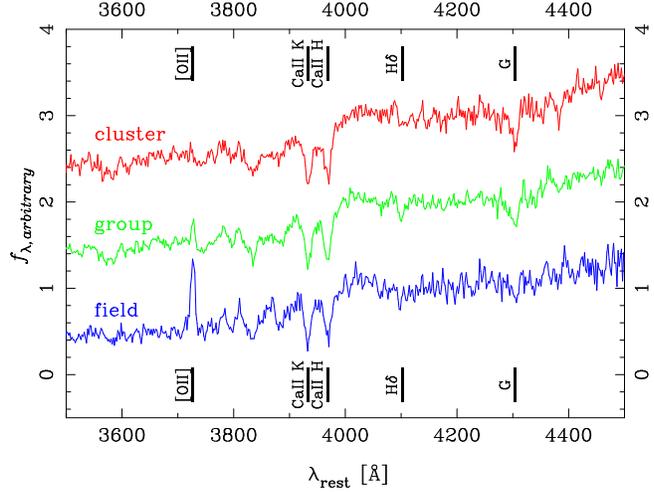}
\end{center}
\caption{
The restframe composite spectra of cluster, group, and field galaxies.
}
\label{fig:spec_comb}
\end{figure}

Fig. \ref{fig:spec_comb} compares the composite spectra.
The continua of the cluster and the group spectra are very similar
and are typical of red passively evolving galaxies, consistent with the
common photometric properties as discussed above.
However, the group spectrum shows a small amount of residual
star formation as characterized by a weak [OII] emission.
The strength of the emission is estimated to be EW[OII$]=4\rm\AA$.
Furthermore, a stronger H$\delta$ absorption is seen in the group spectrum
than in the cluster spectrum.
The field spectrum shows an even stronger [OII] emission (EW[OII$]=13\rm\AA$).
Although the continuum of the field spectrum is similar to those of
cluster and group galaxies, differences can be seen,
e.g. at $\lambda_{rest}\sim3850\rm\AA$, which leads to the difference
in the strength of 4000\AA\, break (section \ref{sec:spec_diag}).

%---------------------------
\subsection{Spectral Diagnostics}
\label{sec:spec_diag}

%---------------------------
\subsubsection{Observational Data}
We now quantify the differences in the composite spectra by measuring
the strengths of 4000$\rm\AA$ break ($D_{4000}$) and H$\delta$ absorption.
The definition of $D_{4000}$ is taken from \citet{bruzual83}.
H$\delta$ absorption lines echo star formation activities that ended
$0.1-1$ Gyr prior to the observed epoch.
We adopt the H$\delta_F$ index defined by \citet{worthey97}.
This index uses a narrower window than that of H$\delta_A$.
Since our spectra are not of very high $S/N$, a narrower window that excludes
the tails of absorption line features is better suited.
We take the median fluxes in the continuum windows to determine
a pseudo-continuum level.
Appendix \ref{sec:concerns} discusses possible sources of uncertainties
in our measurements in detail.

We measure $D_{4000}$ and H$\delta_F$ from the composite spectra of
the cluster/group/field red galaxies at $z\sim0.8$.
In deriving H$\delta_F$, we correct for an H$\delta$ emission
from gaseous nebulae.
We evaluate a typical amount of the H$\delta$ emission filling
from the galaxies at $z=0$ (SDSS) that have identical $D_{4000}$
and EW[OII] to those of the $z\sim0.8$ sample.
We then correct the H$\delta_F$ index for the emission filling.
Note that transforming an [OII] flux into an H$\alpha$ flux, then into
an H$\delta$ flux gives a consistent amount of the emission filling
according to the SDSS data.

A measurement error is estimated as follows.
We perform the bootstrap resampling of the input spectra,
and a composite spectra is generated in each resampling.
Then $D_{4000}$ and H$\delta_F$ are calculated.
This procedure is repeated 5,000 times,
and a 68 percentile interval of the distribution is quoted as an error.
Note that it is difficult to quantify an error in the correction
for the H$\delta$ emission filling, and its error is not included
in the error quoted below.

We utilize the SDSS data for a local counterpart of our $z\sim0.8$ data.
A spectroscopic sample is drawn from the public DR2 \citep{abazajian04}.
For details of the SDSS data,
refer to \citet{stoughton02} and \citet{strauss02}.
For a fair comparison with galaxies $z\sim0.8$, we select the SDSS galaxies
with $M_V\lesssim M_V^*+1$ from the \citet{tanaka05} catalogue
(Main Galaxies at $0.005<z<0.065$).
Note that this magnitude cut corresponds to $m_{z'}\lesssim22$
at $z\sim0.8$ where our spectroscopic galaxies at $z\sim0.8$ fall.
Note as well that we include both blue galaxies and red galaxies
to show that our models, which we describe later, well reproduce
various star formation histories of galaxies at $z=0$.
The spectra are smoothed to our instrumental resolution,
and then $D_{4000}$ and H$\delta_F$ are measured.
We correct for an H$\delta$ emission from a strength of an H$\alpha$
emission assuming the case-B recombination
(H$\alpha=11$H$\delta$ ; \citealt{osterbrock88}).

Measurement errors are estimated from a Monte-Carlo simulation.
A noise is randomly added at each wavelength based on the error estimated by
the SDSS pipeline, and we re-measure $D_{4000}$ and H$\delta_F$.
This procedure is repeated 10,000 times and a 68 percentile interval of
the distribution is quoted as an error.
The SDSS spectroscopic survey is performed with 3 arcseconds diameter fiberes,
which are much smaller than a typical size of galaxies in our $z=0$ sample \citep{tanaka04}.
We argue in Appendix \ref{sec:aperture_bias} that our conclusions are robust
to this aperture bias.

A fundamental question here is the evolutional and environmental connection
between galaxies at $z\sim0.8$ and those at $z=0$.
Do the field red galaxies at $z\sim0.8$ evolve to group/cluster galaxies at $z=0$?
Or are they still in the field at $z=0$?
Fortunately, we find no convincing evidence that the $D_{4000}$ and H$\delta_F$
indices of the $z=0$ galaxies depends on environment --
red field/group/cluster galaxies all have similar $D_{4000}$ and H$\delta_F$
distribution.
That is, the red galaxies at $z\sim0.8$ should evolve to, as shown below,
$D_{4000}\sim2.3$ and H$\delta_F\sim0$ at $z=0$ regardless of environment.
We therefore do not separate the $z=0$ galaxies into various environments.

We present in Fig. \ref{fig:hd_d4000} the distribution of $D_{4000}$ and
H$\delta_F$ indices of galaxies at $z\sim0.8$ (large symbols)
and $z=0$ (contours).
For $z\sim0.8$ galaxies, the $D_{4000}$ indices of cluster and group galaxies
agree within the errors.
This is consistent with the fact that the colours of group and
cluster red galaxies are indistinguishable (see section \ref{sec:composite_spectra}).
Interestingly, there is a hint that the H$\delta_F$ absorption is
stronger for group red galaxies than for the cluster one.
This may suggest that the group galaxies have had recent star formation
at $0.1-1$ Gyr prior to the observed epoch.
The field red galaxies show a smaller $D_{4000}$, and a stronger
H$\delta_F$ absorption, suggesting their younger ages compared to
the red galaxies in denser regions.

Galaxies at $z=0$ show a well-defined sequence on the H$\delta_F$-$D_{4000}$ diagram.
Blue galaxies (small $D_{4000}$) tend to show a strong H$\delta$ absorption,
indicating their active star formation over the past $\sim1$ Gyr.
Red galaxies (large $D_{4000}$) are strongly peaked at
$D_{4000}\sim2.3$ and H$\delta_F\sim0$.
Their weak H$\delta$ absorption suggests that they have not actively formed
stars for a long time ($\gtrsim1$ Gyr).

We compare the $D_{4000}$ and H$\delta_F$ indices of galaxies at $z\sim0.8$
with those at $z=0$.
Red galaxies at $z\sim0.8$ and those at $z=0$ show quite different
distribution in their $D_{4000}$ and H$\delta_F$.
Galaxies at $z\sim0.8$ show a smaller $D_{4000}$ and a stronger H$\delta_F$.
This is qualitatively consistent with the passive evolution model, which
predicts bluer colours (i.e. smaller $D_{4000}$) and a stronger
H$\delta$ absorption at higher redshifts.
However, galaxies at $z\sim0.8$ tend to show a strong H$\delta_F$ for their
$D_{4000}$ compared to local galaxies (group and field galaxies, in particular).
This may suggest that a mode of star formation of group and field red
galaxies at $z\sim0.8$ is different from that of normal star forming galaxies
at $z=0$.
We discuss these trends quantitatively in the next subsection.

%--------------------------------
\subsubsection{Model Predictions}

We compare the observed $D_{4000}$ and H$\delta_F$ with those from
the population synthesis model by \citet{bruzual03}.
As a fiducial parameter set, we adopt the following:
Padova 1994 evolutionary tracks, Chabrier initial mass function between
$0.1-100\rm M_\odot$, solar metallicity, and no dust extinction
(see \citealt{bruzual03} and references therein).
Here we consider three star formation histories to represent the
star formation histories of red galaxies.
One is an instantaneous starburst model, and another is exponentially
decaying model with a decay time scale of $\tau=1$ and 2 Gyr.
The other is a constant star formation model with a starburst at $t=4$ Gyr
since the onset of star formation,
and the star formation is truncated immediately after the burst.
We consider the three starburst strengths; stars that are newly born in
the burst amount to 0\%, 10\%, and 100\% of the mass of the existing stars.
The case for 0\% burst means that star formation is truncated without a burst.
We refer to the above three star formation histories as ``SSP'', ``tau'',
and ``burst'' models, respectively.
Model spectra are smoothed to our instrumental resolution and spectral features
are measured.
Note that our aim here is not to explore a comprehensive range of star
formation histories with many parameter sets such as various extinctions
and metallicities.
However, our simple models should represent the typical paths of 
evolution in $D_{4000}$ and H$\delta_F$.

\begin{figure}
\begin{center}
\leavevmode
\epsfxsize 0.9\hsize \epsfbox{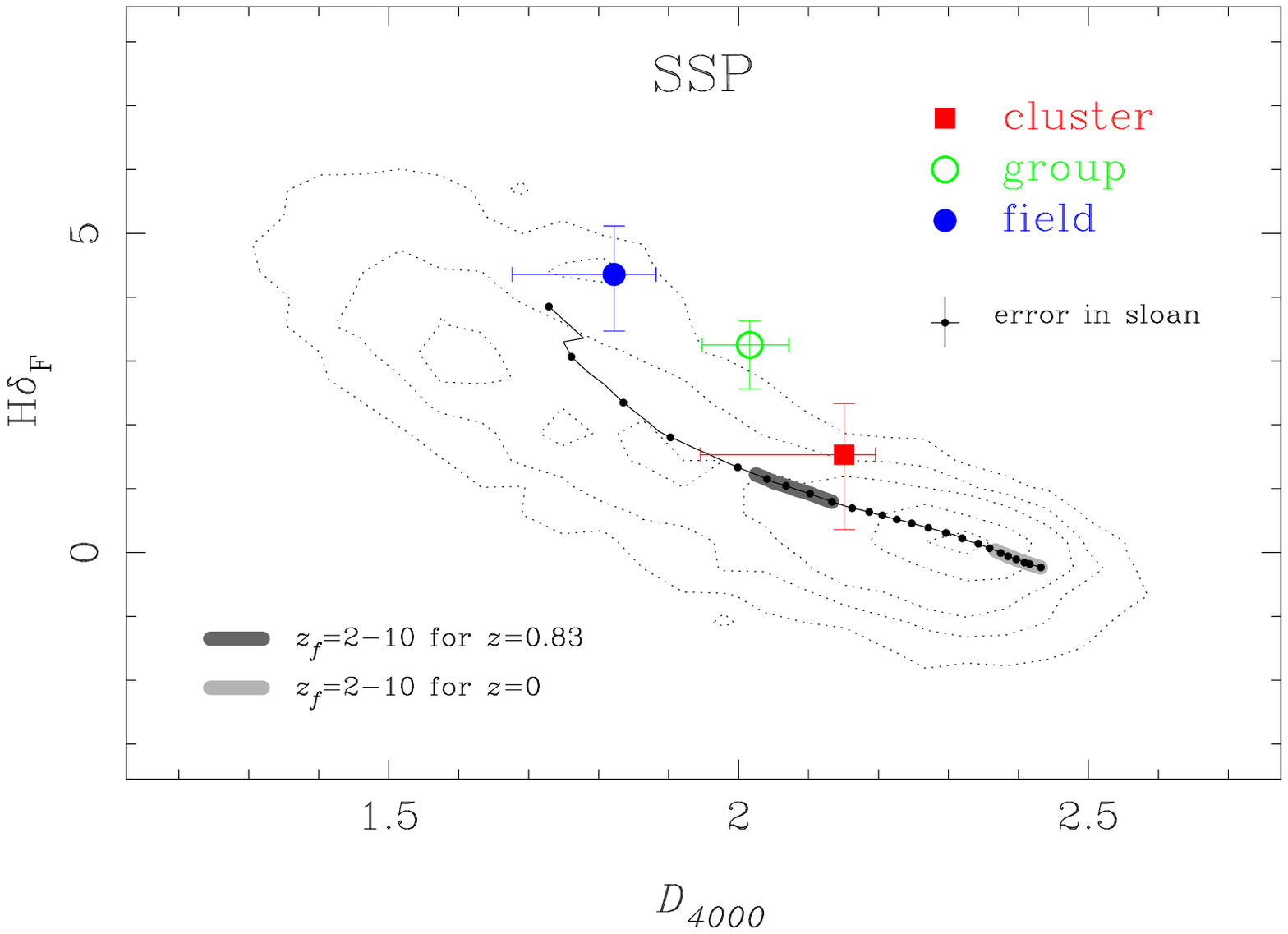}\vspace{0.2cm}\\
\epsfxsize 0.9\hsize \epsfbox{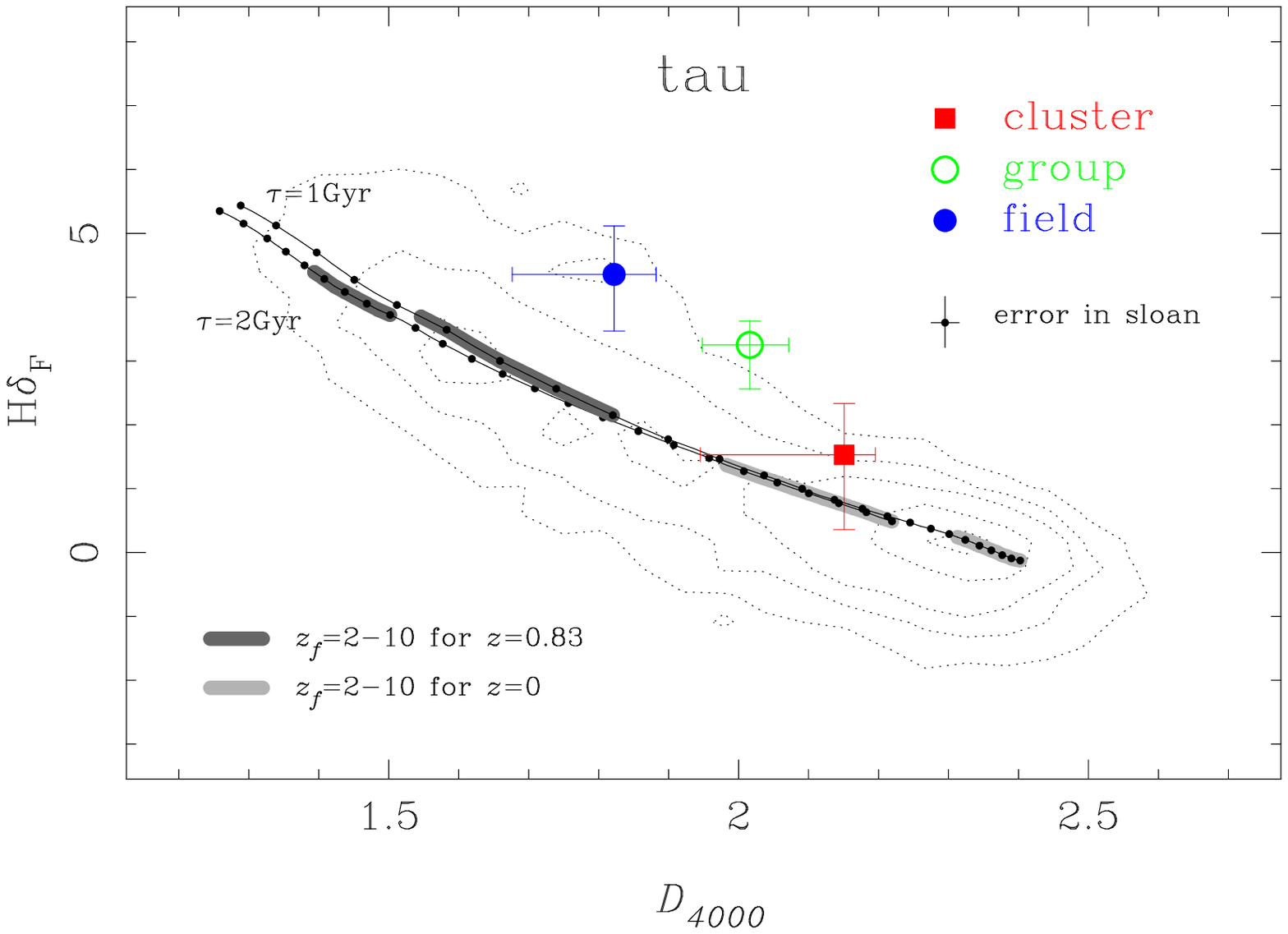}\vspace{0.2cm}\\
\epsfxsize 0.9\hsize \epsfbox{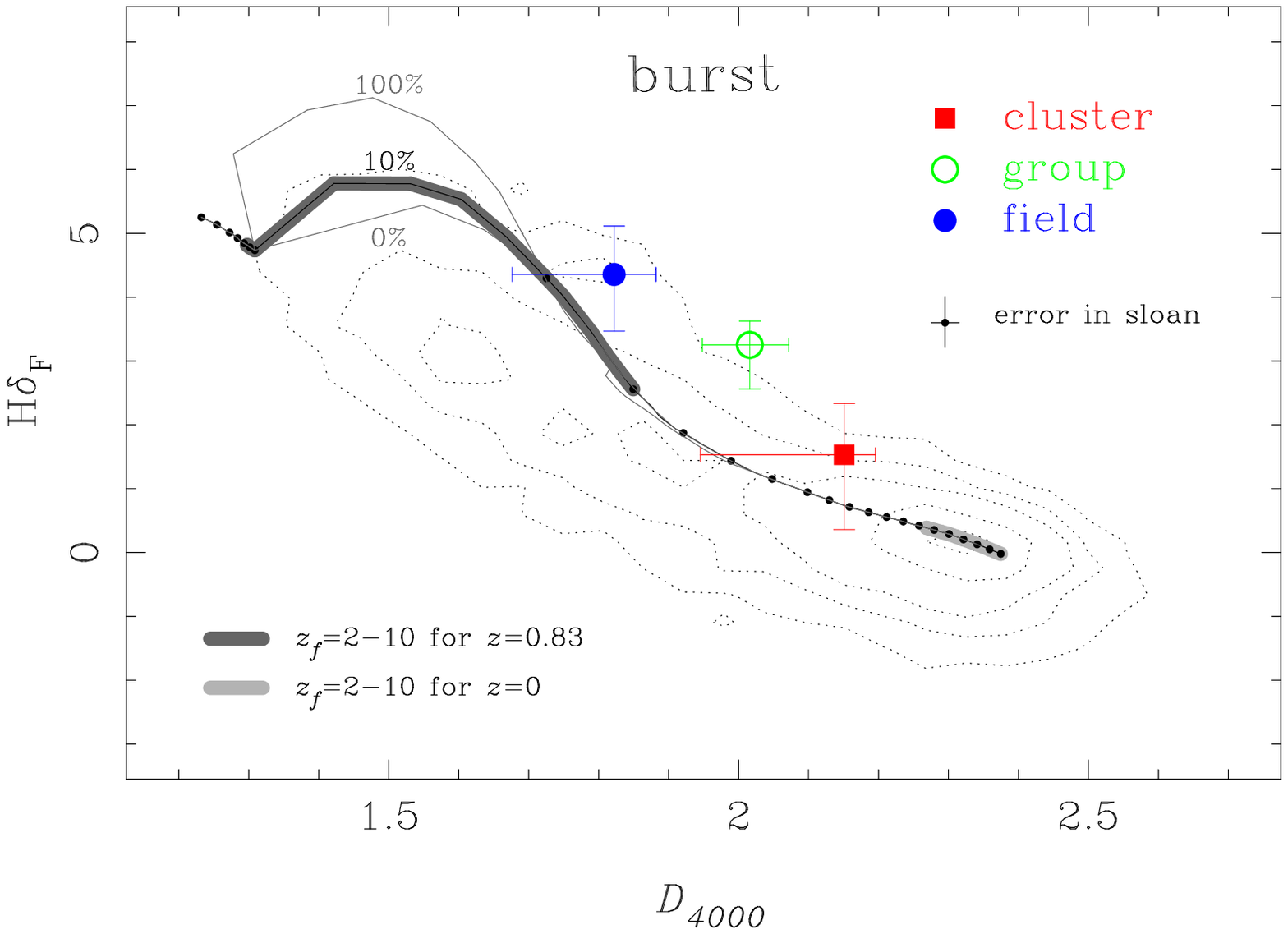}\vspace{0.2cm}
\end{center}
\caption{
The H$\delta_F$ index plotted against $D_{4000}$.
The filled square, open circle, and filled circle respectively show cluster,
group, and field composites at $z\sim0.8$.
The contours show distribution of $z=0$ galaxies brighter than $M_V^*+1$.
The contours enclose 5, 20, 50, 80, and 95\% of the galaxies.
A typical measurement error for $z=0$ galaxies is indicated in each panel.
The panels show different BC03 model tracks; from top to bottom,
SSP, tau, and burst models.
The model predictions are presented as the line-connected points
(points at every 0.5 Gyr).
The model starts from 1 Gyr and ends at 13 Gyr (from left to right).
The regions shaded dark gray and light gray respectively show the model locus
of $z_f=2$ to $z_f=10$ for galaxies at $z=0.83$ and $z=0$, where
$z_f$ is the formation redshift of model galaxies.
Note that, in the burst model, $z_f=2$ and $z_f=10$ for $z=0.83$ galaxies
correspond to 2 Gyr before and 1 Gyr after the burst.
For $z=0$ galaxies, they correspond to 6 and 9 Gyr after the burst.
}
\label{fig:hd_d4000}
\end{figure}

We now put the arguments in the previous subsection on a more
quantitative ground with a help of the model predictions.
As shown in Fig. \ref{fig:hd_d4000}, the various models reasonably
cover the distribution of the observed $z=0$ galaxies.
At $z\sim0.8$, the cluster red galaxies are on the passive evolution model
within the error, and they would naturally evolve into the red galaxies
at $z=0$.
However, the group and the field red galaxies cannot be reproduced
by the simple evolution models such as ``SSP'' or ``tau'' models.
Their H$\delta$ absorptions are too strong for their $D_{4000}$.
We find that SDSS galaxies with EW[OII]$=4\rm\AA$ and EW[OII]$=13\rm\AA$ typically have
$-0.8<{\rm H}\delta_F<0.8$ and $-0.6<{\rm H}\delta_F<2.6$, respectively
(group/field composite spectra show EW[OII]$=4\rm\AA/13\AA$, respectively).
Therefore, the H$\delta_F$ indices of group and field galaxies are clearly
larger compared to normal star forming galaxies.
Field red galaxies could be fit by a burst+sharp truncation model.
However,  they still show a sign of star formation and their
star formation is not completely quenched yet.
It is interesting to note that 0\%, 30\%, and 80\% of individual
cluster, group, and field red galaxies show [OII] emissions.

The group red galaxies cannot be reproduced by any of our model.
In fact, few local galaxies populate around the group galaxies on the
H$\delta_F$--$D_{4000}$ plane.
One may suspect that the correction for the emission filling is not reliable.
However, the correction applied for the red group galaxies is small
($\Delta$H$\delta_F=0.6$), and whether or not we apply the correction
does not significantly alter the result.
It could be due to the dust extinction which makes their colours red \citep{wolf05},
although we do not attempt to perform detailed modeling of dust extinction
in this paper.
As the ``tau'' model demonstrates, a gradual truncation of star formation
does not enhance the H$\delta_F$ absorption and fail to reproduce the
group red galaxies.
The observed strong H$\delta$ absorption favours a scenario that
the star formation is suppressed in a short time scale rather than a slow decline.
We further discuss the physical driver of the truncation of star formation
in groups in section \ref{sec:sfr_truncation}.

%--------------------------figures

%
%
%------------------------------------------------------------DISCUSSION
\section{Discussion}
\label{sec:discussion}

%---------------------------------
\subsection{Comparisons with Previous Studies}
\label{sec:comparison}

\citet{demarco05} reported that the colours of RXJ0153 cluster red galaxies
are consistent with the passive evolution model with $z_f\sim1.9$.
We find that our result is consistent with theirs
(see the ``SSP'' plot in Fig. \ref{fig:hd_d4000}).
They also reported that red [OII] emitting galaxies are preferentially
located in the outskirts of the cluster.
In our analyses, we find that many red [OII] emitting galaxies
in groups and field, and none of cluster red galaxies show [OII] emissions.
Therefore, red [OII] emitters tend to avoid the dense cluster core 
in consistent with Demarco et al.'s result.

\citet{jorgensen05} performed a detailed analysis of absorption lines
of galaxies in the RXJ0153 cluster.
Their results are somewhat different from ours -- they suggested that
the $D_{4000}$ does not evolve from $z=0.83$ to $z=0$, although
they also reported that the evolution of $B$-band luminosity and
high order Balmer lines (H$\delta$ and H$\gamma$) are consistent
with the passive evolution since $z_f\sim4$.
We find that this difference in the evolution of $D_{4000}$ comes
from a difference in $z=0$ data.
There is no significant difference between our $D_{4000}$ measurements
and their $D_{4000}$ measurements for the $z=0.83$ sample,
$D_{4000,z=0.83}\sim2.1$.
However, for red galaxies at $z=0$, we obtain $D_{4000,z=0}\sim2.3$
from the SDSS data, while they obtained $D_{4000,z=0}\sim2.1-2.2$
from their own Perseus and Abell 194 data.
We note that the value $D_{4000,z=0}\sim2.1-2.2$ seems to be inconsistent
with model predictions for evolved red galaxies at $z=0$.
However, we cannot conclusively figure out the cause of
the difference in $D_{4000}$ in the $z=0$ samples at this stage.
We note that \citet{jorgensen05} also reported that some scaling relations
for metal indices, which we do not explore in this paper,
are also inconsistent with the passive evolution.

%---------------------------------
\subsection{Epoch of the Truncation of Star Formation Activities}
\label{sec:sfr_truncation}

\citet{tanaka05} suggested that the truncation of star formation starts from
massive galaxies and fainter galaxies stop their star formation later,
and galaxies follow 'down-sizing' evolution \citep{cowie96,kodama04}.
They showed that, at $z=0.83$, a clear colour-magnitude relation is seen
down to $M_V^*+4$ in cluster environments.
On the other hand, in group environments, only the bright-end ($M_V<M_V^*+2$)
of the relation is seen.
Faint red galaxies do not form a tight relation.
In field environments, no clear relation is seen even at the bright-end.
They therefore suggested that the down-sizing is likely
to depend on environments -- down-sizing is delayed in low-density
environments.

Our results in the previous section support the picture that the truncation of
star formation is delayed in lower-density environments.
While the colour-magnitude distributions of the cluster and the group red
galaxies are identical,
the group red galaxies show a weak {\sc [OII]} emission with a strong
H$\delta$ absorption, which are not seen in the cluster spectrum.
This means that the group red galaxies must have had
recent star formation activities, while the cluster red galaxies have not
formed stars for a long time.
The truncation of field galaxies would occur even later since the
field red galaxies are still actively forming stars.
It may be reasonable to consider that cluster red galaxies have stopped star
formation well in advance, while group red galaxies are just in the process
of truncation, and the field red galaxies have not even stopped star
formation.

Star formation activities reflect an evolutionary stage of the
colour-magnitude build-up.
The red cluster and group galaxies show no/little star formation activities,
and the bright-end of the colour-magnitude relation in these regions is
already built-up.
On the other hand, field red galaxies show rather active star formation and
they do not form a tight relation.
\citet{tanaka05} reported that the bright-end of the field colour-magnitude
relation is built-up down to $z\sim0$.
We expect that the star formation activities of field red galaxies become
significantly weaker by $z=0$.

What physical mechanism truncates star formation activities?
A hint for this question lies in groups.
Group galaxies form a colour-magnitude relation at the bright-end, but not
at the faint-end.
This should indicate that the bright-end of the relation is built-up recently;
the star formation rates of bright galaxies in groups dropped in a recent past.
A close inspection of group galaxies will therefore provide a clue to the physical
driver of the truncation.

There are at least two possible physical mechanisms that can truncate star
formation activities in groups:
galaxy-galaxy interactions (e.g. \citealt{mihos96}) and strangulation \citep{larson80,balogh00}.
Galaxy-galaxy interactions trigger starbursts and star formation is
truncated after the burst.
Strangulation gradually truncates star formation over $\sim1$ Gyr.
These processes can be differentiated by the H$\delta$ absorption --
starburst+truncation enhances the H$\delta$ absorption after the burst,
while strangulation does not trigger the enhancement.
This is shown by the model predictions in Fig. \ref{fig:hd_d4000}
(strangulation should follow a similar track to the ``tau'' model).
It is interesting to note that a sudden truncation of star formation without burst
(0\% burst) also enhances the H$\delta$ absorption.
Ram-pressure stripping \citep{gunn72,abadi99} will truncate star formation
in a relatively short time scale, and the model suggests ram-pressure stripping
could explain the observed H$\delta$ absorption.
However, ram-pressure stripping is not effective in groups, and this is not
a viable mechanism.
The observed strong H$\delta$ absorption in groups then lends support to the scenario
that star formation is quenched in a short time scale,
though the group spectra cannot be reproduced by our fiducial burst model.
It could be that group galaxies experience dusty starbursts, but
the exploration of the dust effects awaits a larger number of higher-$S/N$ spectra.
It seems that galaxy-galaxy interactions would be a viable mechanism.
Interestingly, we find that a composite spectrum of red galaxies
in the filamentary structures (F2 and F3 in Fig \ref{fig:target_fields})
shows a very similar properties to the group composite spectra,
in the sense that $D_{4000}$ and H$\delta_F$ agree within the errors.
Strangulation cannot be effective in filaments and
this would further support galaxy-galaxy interactions.

Further discriminations between the physical drivers
can be made with the information of morphology.
On one hand, strong interactions destroy a disk structure and transform
spiral galaxies into elliptical galaxies.
On the other hand, strangulation do not strongly disturb a disk structure,
and it would transform spiral galaxies into S0 galaxies.
Therefore, the dominant population in groups will give an independent clue to
the physical driver of the truncation of star formation.
Although the mergers between red galaxies may erase this important
dynamical footprint \citep{faber05,tran05,vandokkum05},
the morphological information is still essential in order to
identify the physical driver more conclusively.

In this paper, we confirm the environmental dependence of star formation of
bright red galaxies at $\sim0.8$.
However, the down-sizing phenomenon remains unexplored.
To spectroscopically confirm the down-sizing, we should go fainter than $M_V^*+2$.
To reach such faint magnitudes is difficult even with an 8-m telescope.
In this paper, we reach only $M_V^*+1$ and a number of spectra in each environment is still small.
For these reasons, we do not attempt to discuss the down-sizing at this bright magnitude range.
We expect that faint red galaxies show more active star formation
than bright red galaxies in the same environment.
That is, an epoch of star formation truncation of faint galaxies comes later
than that of bright galaxies.
Another forecast is that bright red galaxies in groups show similar spectra
to faint red galaxies in clusters due to the environmental dependence of
the down-sizing effect.
Data from intensive spectroscopic observations will be able to address these points.

Finally, we recall that our results are based only on one cluster field
at $z\sim0.8$.
A larger sample is essential to step forward.
A typical $S/N$ of the individual spectrum presented in this paper is low,
and the spectra may be affected by systematic uncertainties.
A larger number of higher $S/N$ spectra are clearly needed to
confirm our results and to further explore
star formation histories of galaxies with various metallicity and dust extinctions
as a function of environment and address the down-sizing effect.

%
%
%------------------------------------------------------------CONCLUSIONS
\section{Summary and Conclusions}
\label{sec:conclusion}

We conducted a spectroscopic observation of RXJ0153 at $z=0.83$ with FOCAS on Subaru,
and we obtained 161 secure redshifts.
By combining with redshifts from \citet{demarco05} and \citet{jorgensen05},
we constructed a spectroscopic sample of 347 galaxies.
Our primary conclusions can be summarized as follows.
Note that conclusions 1 and 3 are drawn from our own sample,
and conslusion 2 is from the whole sample.\\

\noindent
1 -- We spectroscopically confirmed the large-scale structures at $z\sim0.8$, which
were first identified photometrically by \citet{kodama05} based on the panoramic imaging data.
Spectroscopic redshifts show a sharp spike at $0.82<z<0.85$,
which is much narrower than the redshift width used for our photometric
selection of member candidates.
This is strong evidence for large-scale structures associated to the
central cluster.
There seems to be two large sheets of galaxies: one at $z\sim0.837$ hosting
the main body of the cluster extending from North to South, and the other
at $z=0.844$ extending from NE to SSW.  They are likely in the process
of interaction.\\

\noindent
2 -- The accuracy of our photometric redshifts depends on galaxy SEDs.
Photometric redshifts are fairly accurate for red galaxies, while
they are less accurate for blue galaxies.
We therefore have to be careful when we discuss a fraction of red/blue
galaxies (e.g. Butcher-Oemler effect) based on phot-$z$.\\

\noindent
3 --
We infer star formation histories of $z\sim0.8$ red galaxies
in the field, group, and cluster environments.
We quantify the $D_{4000}$ and H$\delta_F$ indices,
and compare them with those of SDSS galaxies at $z=0$ and
those from model predictions.
Red cluster galaxies at $z\sim0.8$ have not formed stars for
$>1$ Gyr in the past,
and the passive evolution can naturally link them to present-day red galaxies.
Red galaxies in groups and in the field at $z\sim0.8$ show evidence for
recent activities of star formation as indicated by their relatively
strong H$\delta$ absorptions.
Our current data seem to favour a scenario that star formation is quenched
in a short time scale in groups, and galaxy-galaxy interactions would be
a viable mechanism.
However, higher $S/N$ data are needed to confirm this.

%
%
%------------------------------------------------------------ACKNOWLEDGEMENTS
\section*{Acknowledgments}
We thank Kentaro Aoki, Takako Hoshi, Nobunari Kashikawa, and Youichi Oyama
for their help during the FOCAS observation and data reduction,
and Yoshihiko Yamada for helpful discussions.
We thank the anonymous referee for reviewing the paper.
M.T. acknowledges support from the Japan Society for Promotion of Science (JSPS)
through JSPS research fellowships for Young Scientists.
This work was financially supported in part by a Grant-in-Aid for the
Scientific Research (No.\, 15740126, 16540223) by the Japanese Ministry of Education,
Culture, Sports and Science.
This study is based on data collected at Subaru Telescope, which is operated by
the National Astronomical Observatory of Japan. 

Funding for the creation and distribution of the SDSS Archive has been
provided by the Alfred P. Sloan Foundation, the Participating Institutions,
the National Aeronautics and Space Administration, the National Science Foundation,
the U.S. Department of Energy, the Japanese Monbukagakusho, and the Max Planck Society.
The SDSS Web site is http://www.sdss.org/.
The SDSS is managed by the Astrophysical Research Consortium (ARC) for the Participating
Institutions. The Participating Institutions are The University of Chicago, Fermilab,
the Institute for Advanced Study, the Japan Participation Group, The Johns Hopkins University,
the Korean Scientist Group, Los Alamos National Laboratory, the Max-Planck-Institute
for Astronomy (MPIA), the Max-Planck-Institute for Astrophysics (MPA),
New Mexico State University, University of Pittsburgh, University of Portsmouth,
Princeton University, the United States Naval Observatory, and the University of Washington.

%
%
%------------------------------------------------------------REFERENCES

%
%
%------------------------------------------------------------APPENDIX
\appendix
%-----------------------------------
\section{Uncertainties in the Measurements of the $D_{4000}$ and H$\delta_F$ Indeces}
\label{sec:concerns}

In this Appendix, we discuss possible sources of uncertainties in our
$D_{4000}$ and H$\delta_F$ measurements.
We address effects of the flux calibration, Galactic extinction, and telluric extinction.

\noindent
Flux calibration:\\
No absolute flux calibration is required for the measurements of
$D_{4000}$ and H$\delta_F$, and the question here is how accurate
the relative flux is over the spectral features.
The H$\delta_F$ index adopts a pseudo-continuum around the absorption
feature, and thus a fluxing error should not affect  H$\delta_F$
in a significant way.
As for $D_{4000}$, an accuracy of SDSS flux at the bluest wavelengths
is on the order of a few percent \citep{abazajian04}.
A relative flux is more accurate than this, and hence $D_{4000}$ of
the SDSS galaxies is expected to be accurate to less than a few per cent.
The flux calibration for RXJ0153 should be reliable since
the $D_{4000}$ feature lies around the centre of our spectral wavelength coverage.
We therefore expect the flux error does not affect
our $D_{4000}$ and H$\delta_F$ measurements in a significant way.

\noindent
Galactic extinction:\\
Both our spectra and the SDSS spectra are not corrected for the Galactic extinction.
We estimate E($B-V$)=0.015 for RXJ0153 from the dust map of \citet{schlegel98}.
A typical E($B-V$) of SDSS galaxies is 0.03 to 0.04.
Since the $D_{4000}$ index is measured in a relatively small wavelength window,
the dust extinction affects our $D_{4000}$ measurements by only $\sim1$\%.
We therefore expect our results are robust to the Galactic extinction.

\noindent
Telluric extinction:\\
For $z\sim0.8$ galaxies, strong telluric extinction (A-band)
lies around the spectral features in interest such as H$\delta$.
We have corrected for this extinction (section \ref{sec:spec_obs}),
but we speculate that we have over/under corrected by $\sim$10\%.
The composite spectra are generated from galaxies at slightly different
redshifts, and this uncertainty should be reduced to some extent.
However, it is difficult to evaluate this uncertainty
in our composite spectra.
The $D_{4000}$ and H$\delta_F$ indices measured from the SDSS galaxies
are free from the strong telluric extinctions.

The remaining concern is the effect of systematic uncertainties,
particularly in the $z\sim0.8$ sample.
It is a difficult task to evaluate them.
However, the fact that our $D_{4000}$ measurements are in good agreement
with those by \citet{jorgensen05} suggests that systematic uncertainties
do not dominate the overall error budget.

%-----------------------------------
\section{Aperture Bias}
\label{sec:aperture_bias}

The SDSS spectroscopic survey is performed with the fibre-fed spectrographs.
Each fibre subtends $3''$ on the sky, which is much smaller than a typical size
of galaxies at $z<0.065$ \citep{tanaka04}.
In contrast, a typical extraction aperture adopted in our spectroscopic
data reduction for $z\sim0.8$ galaxies is roughly 1/2 to 1/3 of a size of a galaxy.
Thus, an extraction aperture adopted in the SDSS is smaller compared to
that for $z\sim0.8$ galaxies in a physical size.
This may artificially make differences between $z=0$ and $z\sim0.8$
in the spectral features explored in section \ref{sec:spec_diag}.
We argue in the following, however, our conclusions are robust
to this aperture bias.

In order to quantify the aperture bias, we use spectroscopic data
of nearby galaxies from \citet{jansen00}.
An important feature of the data is that the spectra are obtained for
both the nuclear region and the entire region of a galaxy
('nuclear' and 'integrated' spectra, respectively).
The nuclear spectra typically include 10\% of the light encluded
in the $B_{26}$ isophote, while the integrated spectra include
80\% on average.
Although the extraction aperture adopted in \citet{jansen00} is different
from that of SDSS, comparisons between the nuclear and integrated spectra
will give an estimate of how the aperture bias affects our measurement, at least qualitatively.
The spectra are smoothed to our instrumental resolution, and
$D_{4000}$ and H$\delta_F$ are measured (again the emission filling is corrected).
We do not apply any magnitude cut to the \citet{jansen00} sample to gain the statistics.
Our aim here is not to compare the integrated spectra with our spectra of
$z\sim0.8$ galaxies, but to investigate the aperture bias.

\begin{figure}
\begin{center}
\leavevmode
\epsfxsize 1.0\hsize \epsfbox{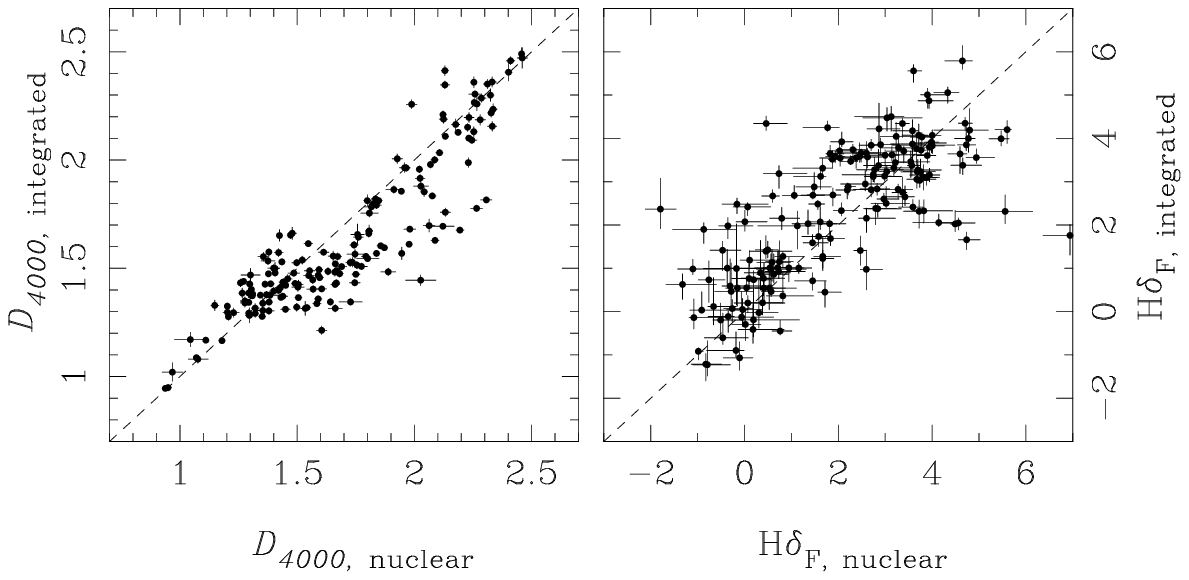}\\\vspace{0.5cm}
\epsfxsize 0.8\hsize \epsfbox{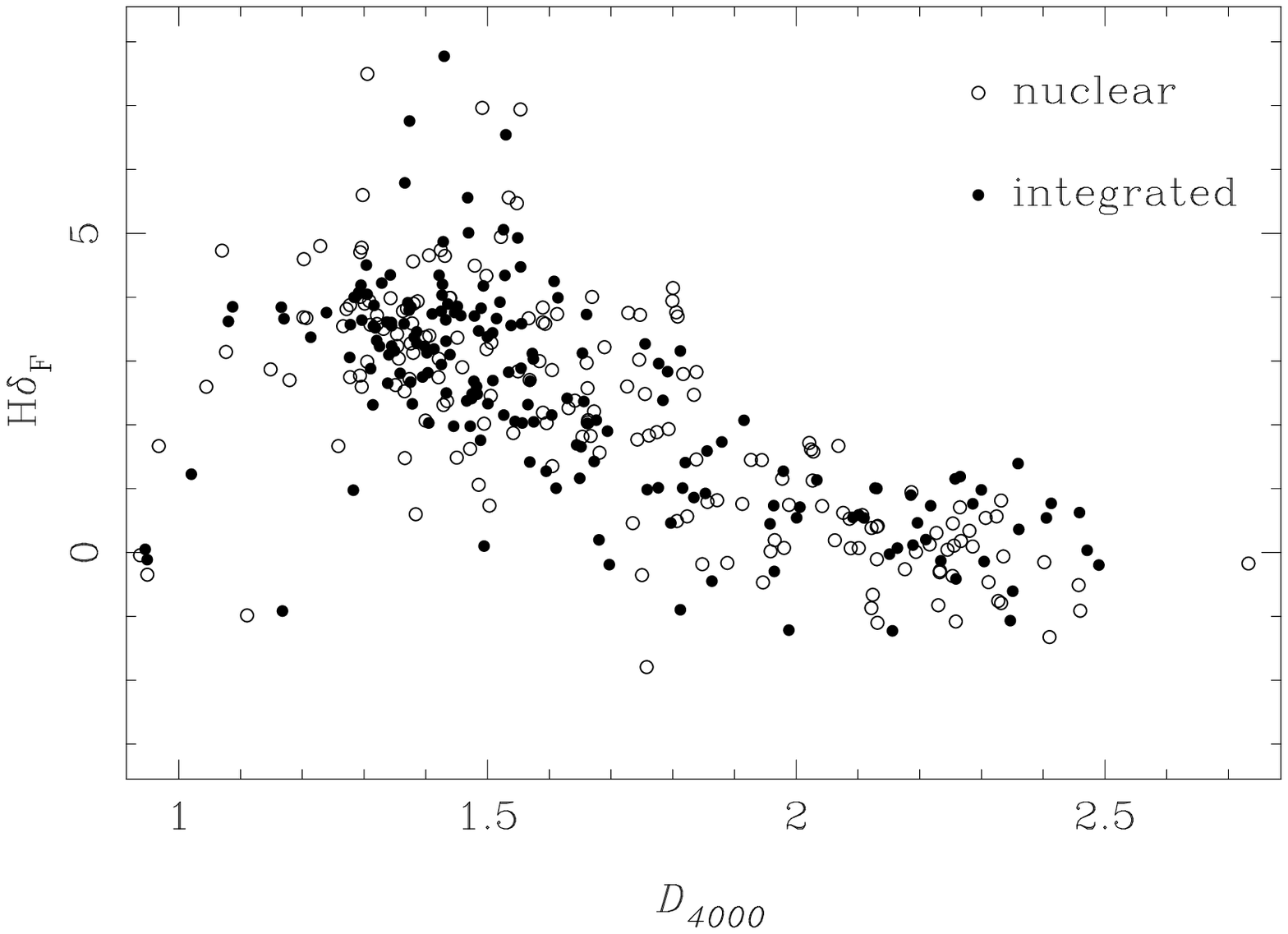}
\end{center}
\caption{
Differences in $D_{4000}$ ({\it Top-Left}) and H$\delta_F$ ({\it Top-Right})
between the nuclear and integrated spectra from \citet{jansen00}.
Distribution of nuclear and integrated spectra on the H$\delta_F$--$D_{4000}$
plane is shown at the bottom.
}
\label{fig:aperture_bias}
\end{figure}

Comparisons between the nuclear and integrated spectra are shown in Fig. \ref{fig:aperture_bias}.
While the overall correlation is encouragingly good, deviations are seen.
The integrated spectra show systematically smaller $D_{4000}$ than the nuclear spectra
at $1.6\lesssim D_{4000,nuclear}\lesssim2.1$.
That is, these galaxies show bluer colours at the outer parts of the galaxies compared to the inner parts.
This is expected since active star formation is often observed in a disk rather than in a bulge.
Very red/blue galaxies do now show such deviation.
As for H$\delta_F$, the correlation between the integrated and nuclear spectra
is relatively good, but a tail toward higher H$\delta_{F, integrated}$ is seen.
Despite these tails, however, the distribution of galaxies on the H$\delta_F$--$D_{4000}$
plane is basically the same for the nuclear and integrated spectra.
Furthermore, for our SDSS sample, spectroscopic fibre typically collects
20\% of the total $g$-band light of a galaxy.
Thus the aperture bias should be smaller than that we see in
Fig. \ref{fig:aperture_bias}.
We therefore conclude that the aperture bias has no significant effect on our conclusions.
Note that the integrated spectra seem to show a smaller scatter on the
$D_{4000}$--H$\delta_F$ plane than the nuclear spectra.
Thus, if the indices are measures for the entire part of galaxies,
the scatter of SDSS galaxies will be smaller than
we observe in Fig. \ref{fig:hd_d4000}.

%--------------------------figures

\end{document}